\documentclass[twocolumn,showpacs,preprintnumbers,amsmath,amssymb,prl]{revtex4-1}

\usepackage{graphicx}% Include figure files
\usepackage{dcolumn}% Align table columns on decimal point
\usepackage{bm}% bold math
\usepackage{hyperref}% add hypertext capabilities
\usepackage{hyperref}
\usepackage{color}
\usepackage{amsmath,amssymb}
\usepackage{times}
\begin{document}

\preprint{Under submission to PRL}
%\preprint{APS/123-QED}

\title{Kolmogorov scaling bridges linear hydrodynamic stability and turbulence}% Force line breaks with \\
%\thanks{A footnote to the article title}%

\author{Stefania Scarsoglio$^1$, Francesca De Santi$^2$ and Daniela Tordella$^{2}$}
\email{daniela.tordella@polito.it}

\affiliation{$^1$Department of Water Engineering, Politecnico di Torino, Torino, Italy\\ $^2$Department of Aeronautics and Space Engineering, Politecnico di Torino, Torino, Italy}%

%\email[]{Your e-mail address}
%\homepage[]{Your web page}
%\thanks{}
%\altaffiliation{}

\date{\today}% It is always \today, today,
             %  but any date may be explicitly specified

\begin{abstract}
The way in which kinetic energy is distributed over the multiplicity of inertial (intermediate) scales is a fundamental feature of turbulence. According to Kolmogorov's 1941 theory, on the basis of a dimensional analysis, the form of the energy spectrum function in this range is the $-5/3$ spectrum. Experimental evidence has accumulated to support this law. Until now, this law has been  considered a distinctive part  of the nonlinear interaction specific to the  turbulence dynamics. We show here that this picture is also present in the linear dynamics of  three-dimensional stable perturbation waves in the intermediate wavenumber range. Through extensive computation of the transient life of these waves, in typical shear flows, we can observe that the residual energy they have when they leave  the transient phase and enter into the final exponential decay shows a spectrum that is very close to the $-5/3$ spectrum. The observation times also show a similar scaling. The scaling depends on the wavenumber only, i.e. it is not sensitive to the inclination of the waves to the basic flow, the shape-symmetry  of the initial condition and the Reynolds number.
\end{abstract}

% PACS, the Physics and Astronomy
                             % Classification Scheme.
%\keywords{Suggested keywords}%Use showkeys class option if keyword
                              %display desired
\maketitle

In fluid systems, stability  and turbulence are the two faces of the same coin. The existence of equilibrium: in one case laminar, and steady in the mean in the other. The link between these two faces is transition \cite{D2002,CJJ2003,SH2001,K41,F95,SA97}.

Unfortunately, or fortunately, depending on the circumstances, turbulence is the rule and not the exception in fluid motion. When the energy forcing in the system is sufficiently high, transition to turbulence occurs in the short or in the long term. In principle, stability and turbulence studies are intimately connected. In practice, the relevant literature is split into two quite distinct fields (however, counterexamples exist, see e.g. \cite{Hof2004,Hof2006}). The main  reason is that  the stability can be defined physically as the ability of a dynamical system to be immune  to small disturbances, which necessarily leads to the linearization of the mathematical formulation. This is something which cannot occur in turbulence where a great number of several interacting scales are always active and experimentally observable.

However, at any instant,  laminar systems host a multiplicity of scales: the small perturbations which randomly enter the system and, in the linear framework, evolve independently from each other. Although linearity on  one hand allows each evolution to be determined singularly, on the other, it should be recalled that  a large number of perturbations (not even bounded, in principle)  are present at the same time. In this work, we have tried to consider and observe the collective behavior of small perturbations, in particular, those filling the intermediate range of wave lengths that the system can host.  The aim is the understanding and discovering of possible similarities with turbulence behavior.

As an example, in order to understand whether, and to what extent, spectral representation can effectively highlight the nonlinear interaction that occurs among different scales, it could be useful to consider the state that precedes the onset of both instability and turbulence in flows.  In this condition,  even if  stable, the system  is however subject to a swarming of small arbitrary three-dimensional  perturbations that constitutes a system
of multiple spatial and temporal scales  subject to all the processes included in the  Navier-Stokes equations: linearized convective transport, linearized vortical stretching and tilting, and  molecular diffusion. If we leave nonlinear interaction of the different scales, the other features are tantamount to the features of the turbulent state.

If it were possible to observe such a system, by computing and comparing a large set of three dimensional waves, and build spectra, it would be possible, among others, to determine if a power  scaling in the intermediate  range exists and, in case, to  compare it with the exponent of the corresponding developed turbulent state (notoriously equal to - 5/3). In the case a power scaling exists, two possible situations can therefore appear. A - The exponent difference is large, and as such, is a quantitative measure of the nonlinear interaction in spectral terms. B - The difference is small. %This would be even more interesting, because
This  would  indicate a higher level of universality on the value of the exponent of the intermediate range (the inertial range in turbulence),  not necessarily associated to the nonlinear interaction.

For this purpose, by solving a large number of initial-value problems, we have determined  a large set of transient solutions (a database of 2400 solutions, see the Supplemental Material, section 5) for two typical shear flows: a plane channel flow, archetype of wall flows, and a two-dimensional bluff-body wake, archetype of free-flows, see Fig. 1 a. Perturbations that randomly arrive in the system undergo a transient evolution which is ruled  by the initial-value problem associated to the Navier-Stokes linearized formulation \cite{K1880,O1907a,O1907b}.
These problems must be parameterized through the principal  physical and geometrical quantities that can influence the life of perturbation waves, the angle of obliquity, the symmetry, the polar wavenumber and Reynolds number. For instance,  the wavelength of the waves can be varied in a range as large as the range of scales that typically fill the field when the  system is in the so-called fully developed  turbulent state.

There exists  many  kinds of transient behavior, very different and not all of which is trivial (for a description, see the text below and in particular the overview presented in figure 2). The transient lives can last a few  basic time scales (external length referred to the velocity scale of the basic flow) as well as order $10^4$ time scales. During these lives energy can be accumulated, then smoothly lost or lost and acquired again. Usually, different inner time scales appear. For instance, the pulsation can change in a discontinuous way before the asymptotic state is reached. %If these transients, obtained in association with arbitrary, statistically selected, initial conditions, could be injected in a statistical way into a temporal observation window, we could obtain  a  close representation of the perturbation state that precedes the onset of instability-turbulence.
This  very rich scenario is met out of any self-interaction and interaction with other waves and, to some degree, is reminiscent, at least qualitatively,  of the  turbulence phenomenology.

So, the question arises: how to compare spectrally the set of very different transients of large, intermediate and short waves? Our answer is that whatever be the difference in the wave lives, a common phase exits: the time interval  where the wave exits the transient and enters the exponential asymptotic state. We thus consider the residual kinetic energy  owned by the waves in this interval and build a spectrum in the wavenumber space.

Let us now describe synthetically the basic flows we have here analyzed, as well as the relevant transient computation. The basic flows  for a channel flow are  represented by the Poiseuille solution (see Fig. 1b and SM, section 4), and for a wake by the first two order terms of the Navier-Stokes asymptotic solution described in \cite{TB03} (see Fig. 1c and SM, section 4). The channel flow is homogeneous along the streamwise and spanwise directions ($x$ and $z$). The  profiles only vary with the coordinate $y$. In the case of the wake, the profiles also slowly evolve with $x$. As a consequence, the flow is not perfectly parallel and we consider two fixed longitudinal stations in the region of the wake where  entrainment is present \cite{TS09}, $x_0 = 10, 50$ ($x_0$ is the distance from the body normalized over the  body length). Different Reynolds numbers (the dimensionless control parameter that gives  the order of magnitude of the ratio  between convection and molecular diffusion) are considered for each example of flow:  subcritical  (steady laminar solution: $Re=30$ for the wake, $Re=500$ for the channel),  supercritical (unstable laminar wake $Re=50, 100$ and  turbulent channel $Re=10000$). An initial-value problem (IVP) for small arbitrary  three-dimensional vorticity perturbations imposed on the basic shear flows is then considered. The viscous
perturbation equations are combined in terms of the vorticity and velocity \cite{CD90}, and are solved by means of a combined Fourier--Fourier (channel) and Laplace--Fourier (wake) transform in the
plane normal to the basic flow, see SM section 4 and \cite{CJJ2003,STC09,STC10}.
%In the case of the channel flow, the dependent variables are  in fact Fourier-Fourier transformed in the $x$ and $z$ directions, which are both field homogeneity directions, while, in the case of the wake flow,  the variables are  Fourier transformed along $z$, the only homogeneous direction of the system, and Laplace transformed along $x$ ($0 < x < \infty)$.
%This treatment allows a simplification to be made of the governing equations, so that it is possible to observe long transients, which can last thousands of time scales. This result would not be possible over an acceptable lapse of time if a direct numerical integration of the linearized Navier--Stokes equations were carried out.
The exploration is conducted  with respect to physical quantities, such as the  polar wavenumber, the angle of obliquity, the symmetry of the perturbation, the flow control parameter, and, for the wake, which is not parallel, the position downstream of the body.
For further details on the formulation and the numerical methods used to solve the initial-value problems see the Supplemental Material (SM, section 4).

\begin{figure}
\begin{center}
\begin{minipage}[b]{0.7\columnwidth}%\begin{tabular}{c c}
\hspace{-1.3cm}
\includegraphics[width=1\columnwidth]{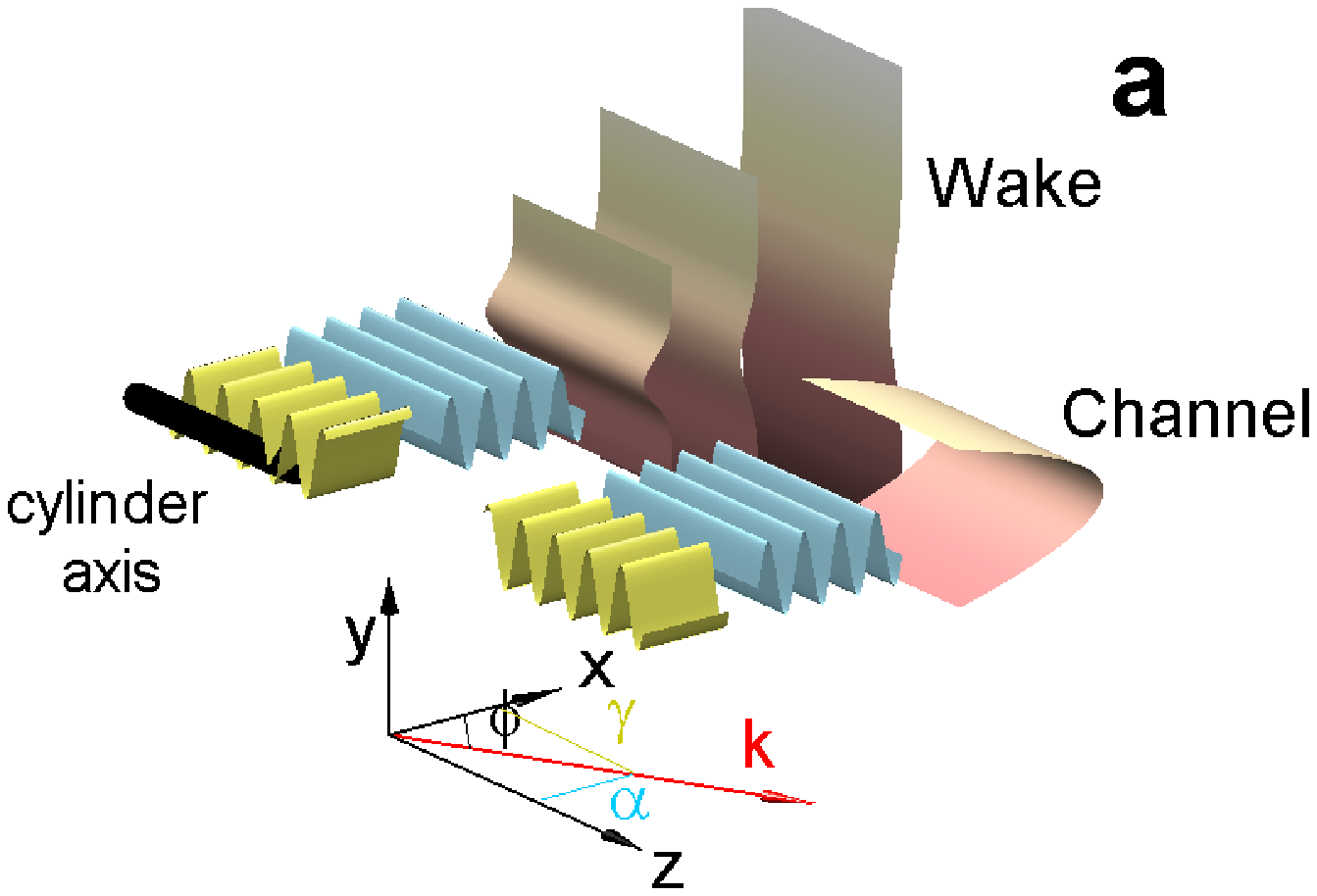}
\end{minipage}
\hspace{-0.5cm}
\begin{minipage}[b]{0.29\columnwidth}
\includegraphics[width=1.2\columnwidth]{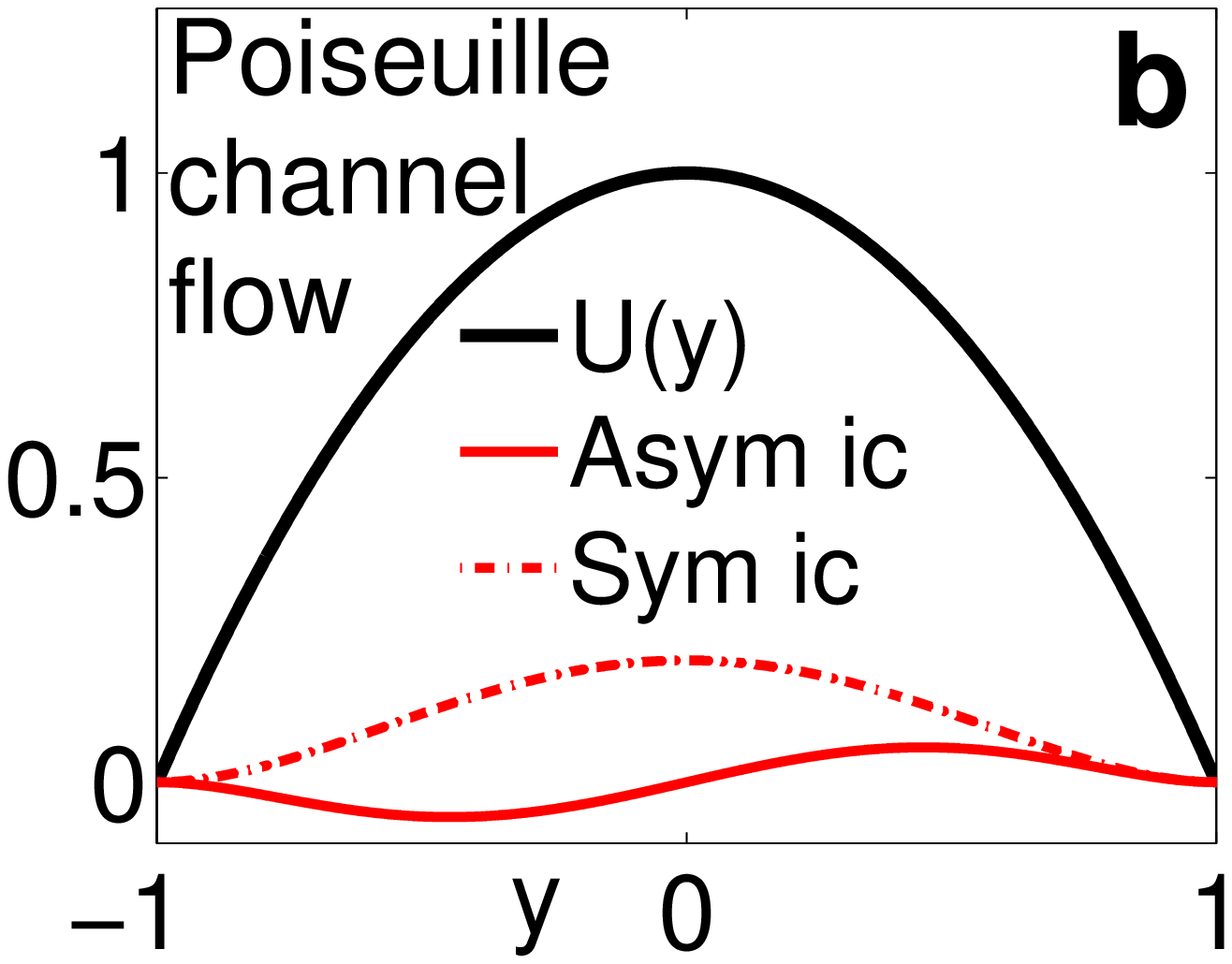}\\
\includegraphics[width=1.2\columnwidth]{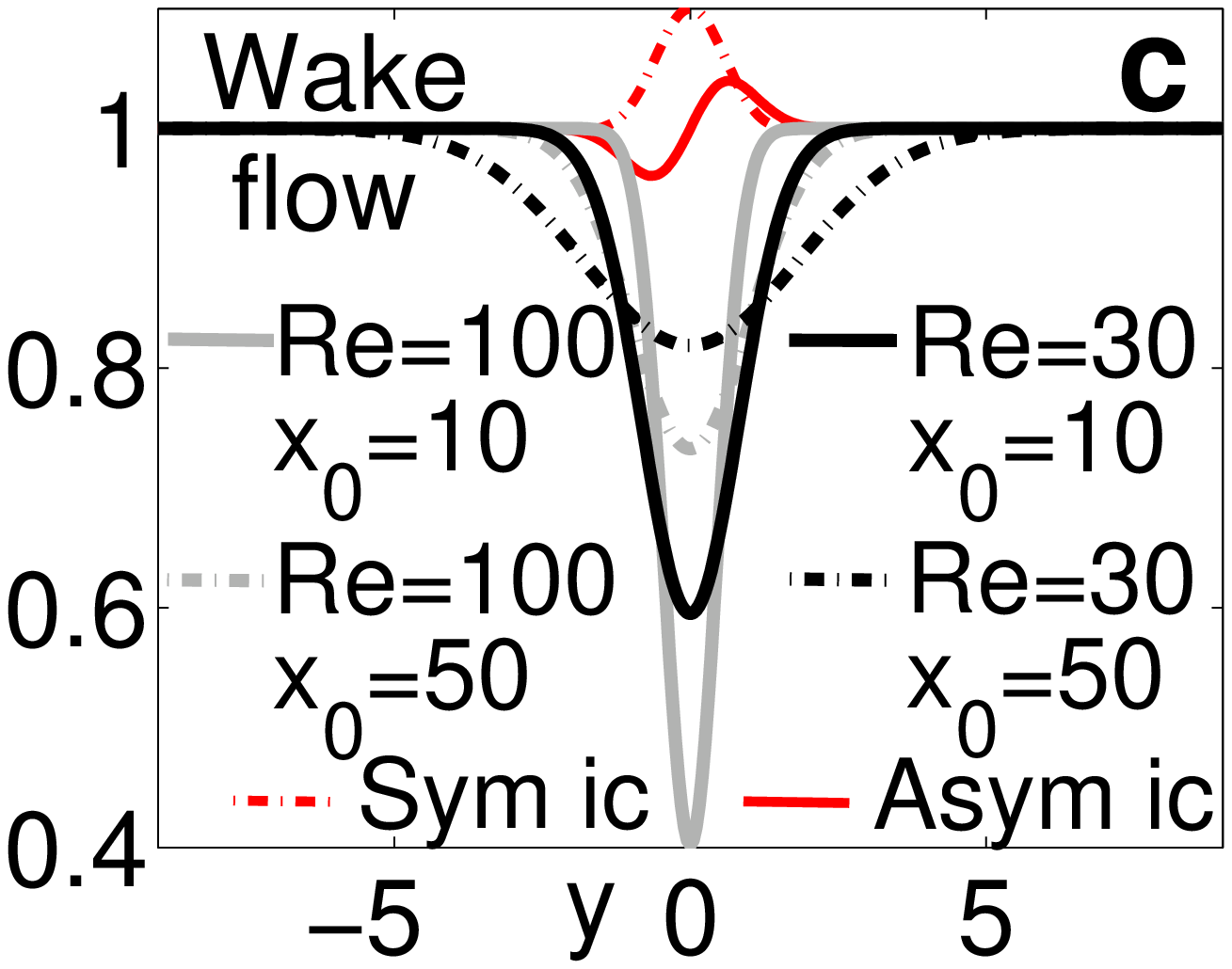}
\end{minipage}
\end{center}
%\vskip -6mm
\caption {(a)  Base flows and perturbation scheme. The flow profiles are qualitatively represented in pink.  The perturbation is represented by the blue and the yellow waves which propagate  in the  $x$ and $z$ directions, respectively. $k = \alpha \textbf{i} + \gamma \textbf{k}$ is the wavenumber, $\phi$ is its inclination angle with respect to the basic flow. (b)-(c) Initial conditions and velocity profiles. The red  dotted and solid lines are the symmetric and asymmetric parts of the initial perturbations, respectively. The grey and black lines in the wake panel (c) schematically  show the velocity variation with the Reynolds number ($Re$) and the position in the wake, $x_0$. Grey lines: supercritical $Re$ (the critical value is $47$), black lines: subcritical value. Dotted lines: far wake field, solid lines: intermediate field.}
\label{fig:vis}
\end{figure}

To measure the temporal evolution of the energy of each perturbation, we define the kinetic energy density, $\displaystyle{e(t; \alpha, \gamma) = \frac{1}{2} \int_{-y_f}^{+y_f} (|\hat{u}|^2
+ |\hat{v}|^2 + |\hat{w}|^2) dy}$, where $-y_f$ and $y_f$ are the computational limits of the domain, $\hat{u}$, $\hat{v}$ and $\hat{w}$ are the transformed velocity components of the perturbation. We can also define the amplification factor, $G$, as the kinetic energy density normalized with respect to its initial value, $G(t; \alpha, \gamma) = e(t; \alpha, \gamma)/e(t=0; \alpha, \gamma)$.

\noindent In terms of amplification factors, the early transient evolution offers very different scenarios for which we present in the following a summary of relevant cases. For example, for both base flows, we have observed that the orthogonal waves are always asymptotically stable. However, the perturbations are able to reach very high maxima of the amplification factor $G$ (of the order of $10^4$, see Fig. 2b) before the transients are extinguished. Non-orthogonal asymmetric waves present as well transients which are not trivial at all (see blue and red solid lines in Fig. 2a and 2c). These perturbations  are slightly amplified in the early stage of their lives, then decrease for several hundreds of time scales and in the end they grow with the same slope of the correspondent symmetric waves (compare the asymptotic trends of non-orthogonal solid and dotted curves in panels a and c of Fig. 2). During the decreasing phase, these transients clearly show an initial oscillatory time scale associated to a modulation in amplitude of the average value of the pulsation in the early transient, and which is different from the asymptotic value of the pulsation (see the insets in panels a-c in Fig. 2)\cite{STC09}. Here the pulsation (angular frequency), $\omega$, is defined as the time derivative of the wave phase at a fixed transversal position (see SM, section 4). Thus, the system exhibits two distinct temporal oscillatory patterns, the first, of transient nature, and, the second one, of asymptotic nature.

\noindent As a general comment, the most important parameters affecting these configurations are the angle of obliquity, the symmetry, and the
polar wavenumber. While the symmetry of the disturbance influences the transient behavior to a great extent and leaves the asymptotic fate unaltered, a variation in the obliquity and in polar wavenumber can significantly change the early trend as well as the final stability configuration.

\begin{figure}
%\vspace{-2cm}
\centering
\begin{minipage}[]{0.48\columnwidth}
\includegraphics[width=\columnwidth]{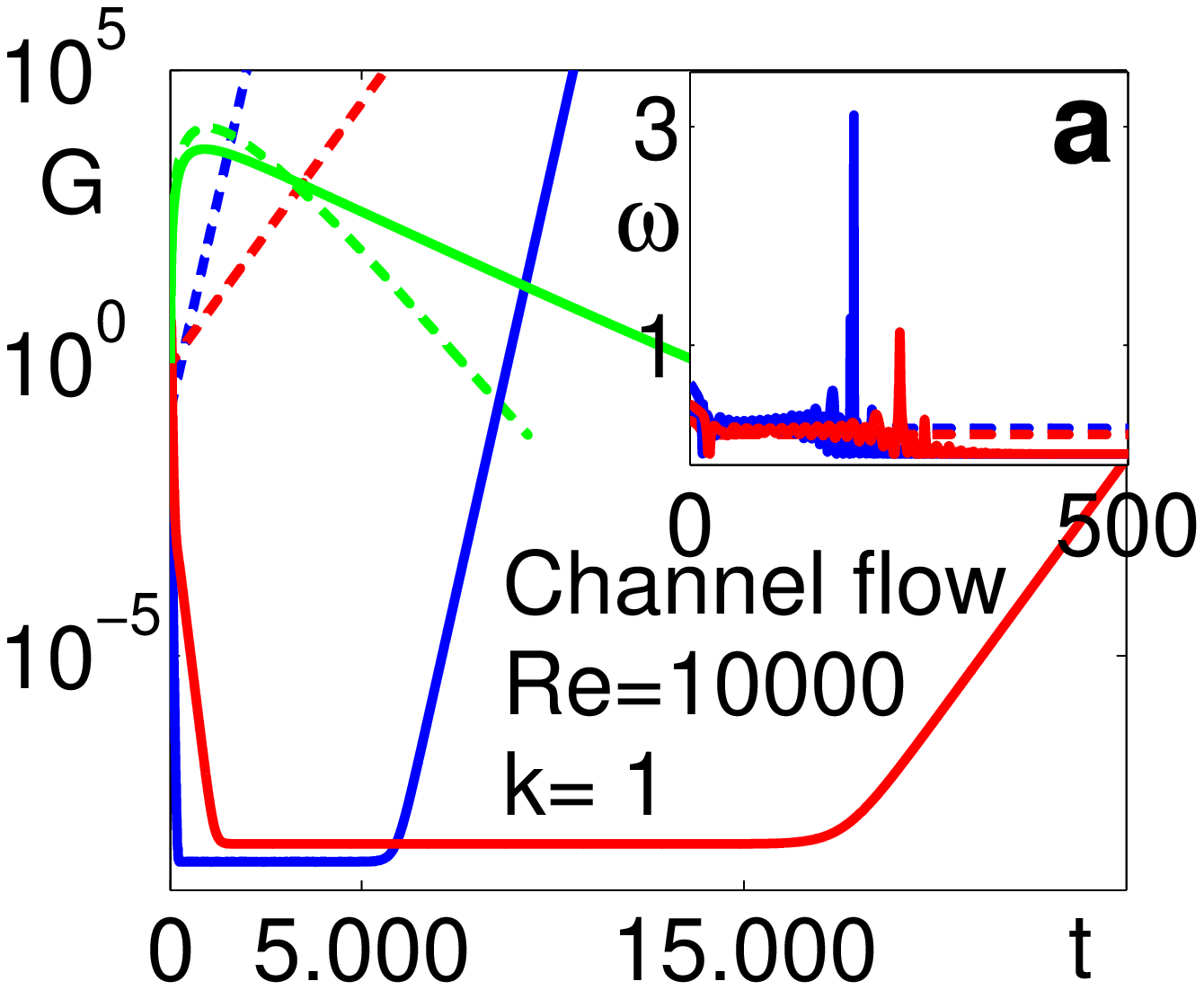}
\end{minipage}
\begin{minipage}[]{0.48\columnwidth}
\includegraphics[width=\columnwidth]{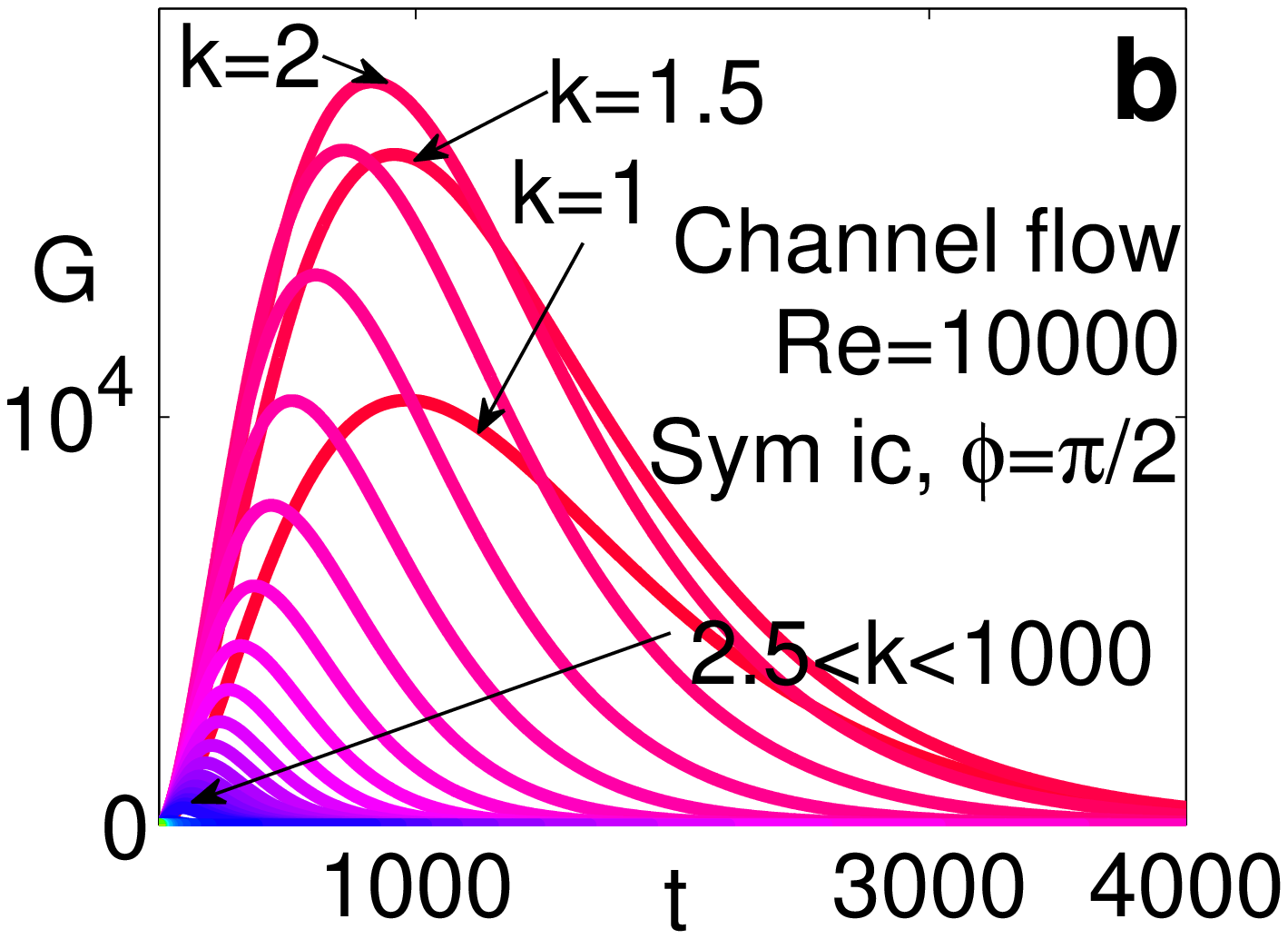}
\end{minipage}
\centering
\begin{minipage}[]{0.48\columnwidth}
\includegraphics[width=\columnwidth]{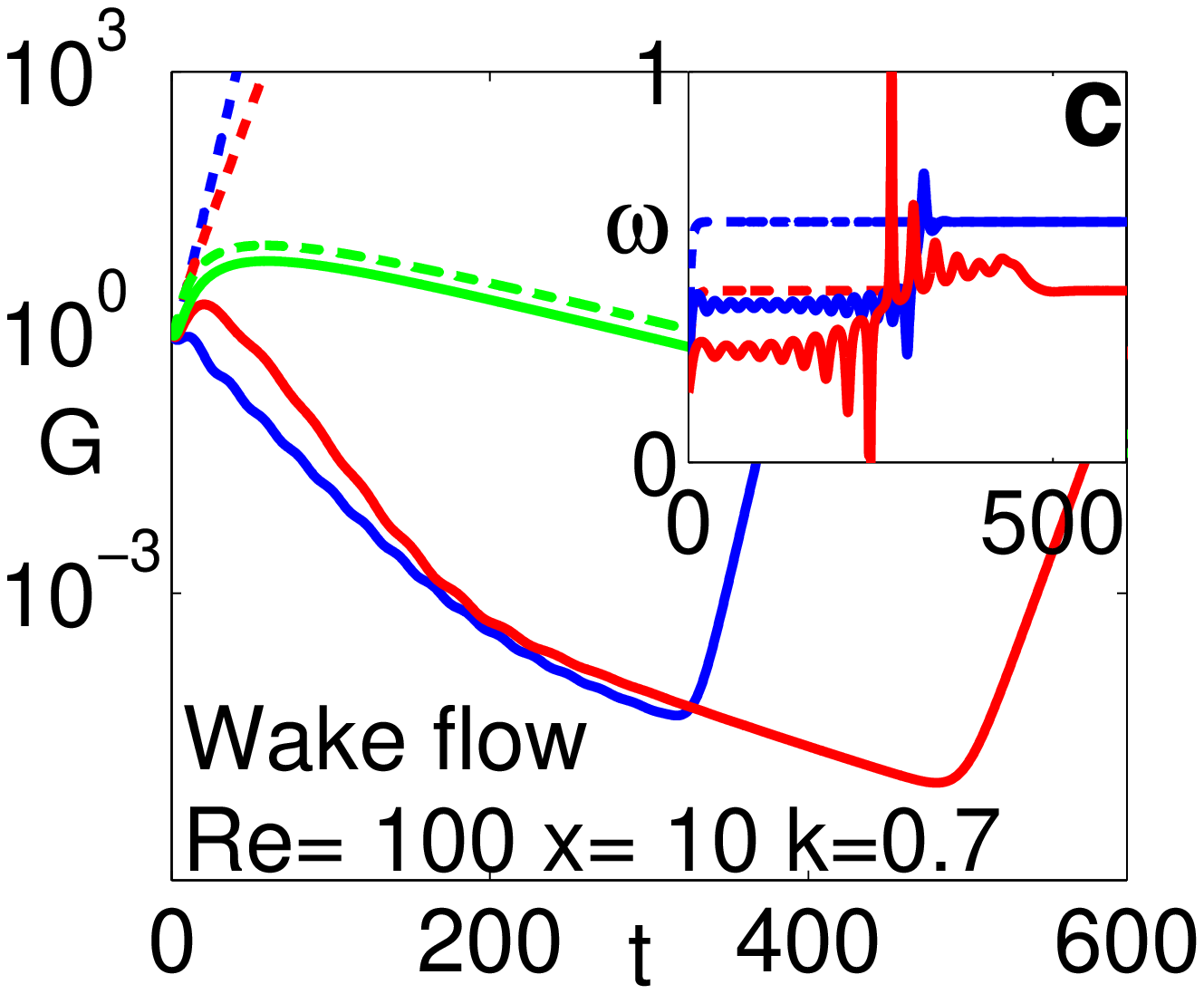}
\end{minipage}
\begin{minipage}[]{0.48\columnwidth}
\includegraphics[width=\columnwidth]{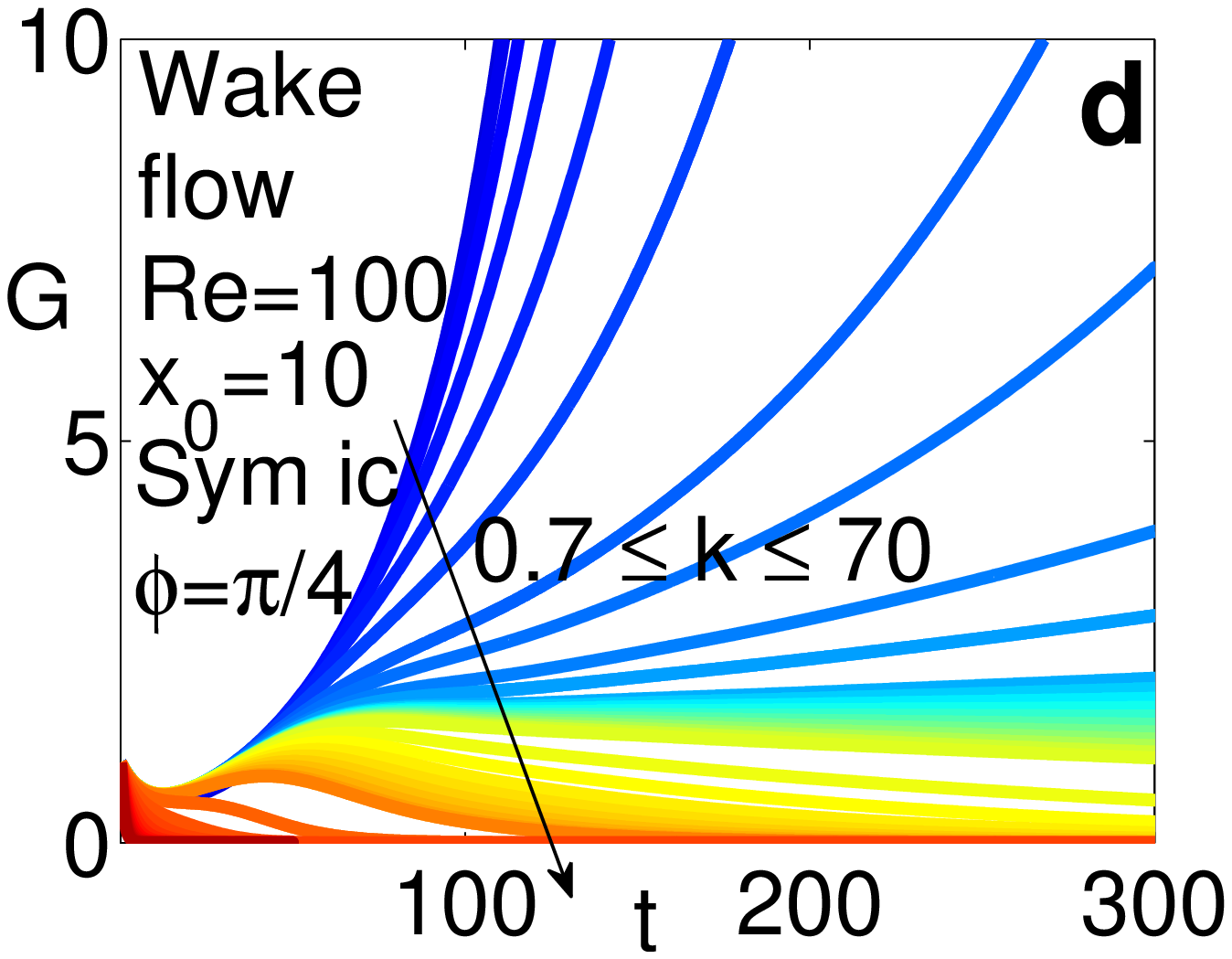}
\end{minipage}
\centering
\begin{minipage}[]{\columnwidth}
	\includegraphics[width=\columnwidth]{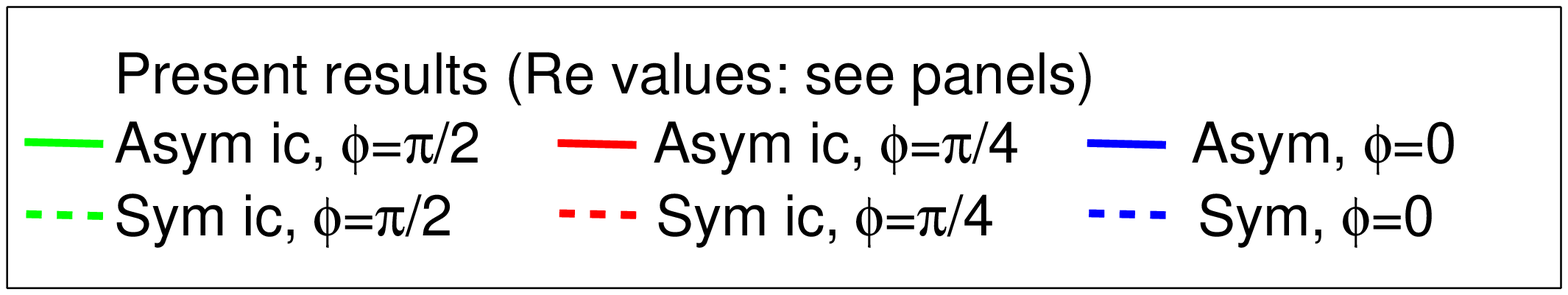}
\end{minipage}
\caption{Collection of transient lives of the perturbations observed through the amplification factor $G$. Left column: the magnitude of the wavenumber is fixed, while the obliquity and symmetry vary (see the legend on the bottom). %The insets highlight the typical formation of time scales others than the flow external time scale and the wave period ($< G >$ in the inset ordinate is the amplification factor averaged over the time interval shown in the inset).
The insets (temporal evolution of the pulsation, $\omega$) highlight the typical formation of time scales others than the flow external time scale and the wave period. Right column: transients variation with changes in the wavenumber magnitude.}
\label{fig:tr1}
\end{figure}

\noindent The asymptotic behavior for the plane wake, for disturbances aligned to the flow,  is shown
to be in excellent agreement \cite{ETC12} with 2D spatio-temporal modal analyses \cite{TSB06,BT06} and with the laboratory determined frequency and wave length of the parallel vortex shedding at $Re = 50$ and $100$ \cite{W89}. See also in the SM, the section Asymptotic pulsation (section 3,  figure S3) where information on the channel flow pulsation measured in the laboratory and from the present IVP computations are given.
%It should be noted that the agreement between the IVP results and the normal mode theory is not obtained using the most unstable wave, given by the Orr-Sommerfeld dispersion relation at any section of the wake, as the initial condition, but arbitrary initial conditions.

%The solutions gathered in Fig.2 reveal the existence of  many  kinds of transient behaviour, not all of which is trivial. If these transients, obtained in association with arbitrary, statistically selected, initial conditions, could be injected in a statistical way into a temporal observation window, we could obtain  a  close representation of the perturbation state that precedes the onset of instability-turbulence. This  very rich scenario is reminiscent, at least qualitatively,  of the  turbulence phenomenology.

We now come back to the spectral analysis of common phase in the lives of the perturbations, that is the transition between the end of the transients and the settlement of the asymptotic  condition. To compare the residual kinetic energy of the waves in correspondence to this transition, we assumed that when the  asymptotic exponential temporal behavior is reached, the temporal growth rate, $r$, defined as $r(t; \alpha, \gamma) = log(e)/(2t)$ \cite{CJJ2003}, must  approach a real constant.

%\begin{equation}
%r(t; \alpha, \gamma) = \frac{\displaystyle{\frac{dG}{dt}}}{G}. \label{IVP2_tgr}
%\end{equation}

\begin{figure}
\centering
\begin{minipage}[]{0.48\columnwidth}
	\includegraphics[width=1\columnwidth]{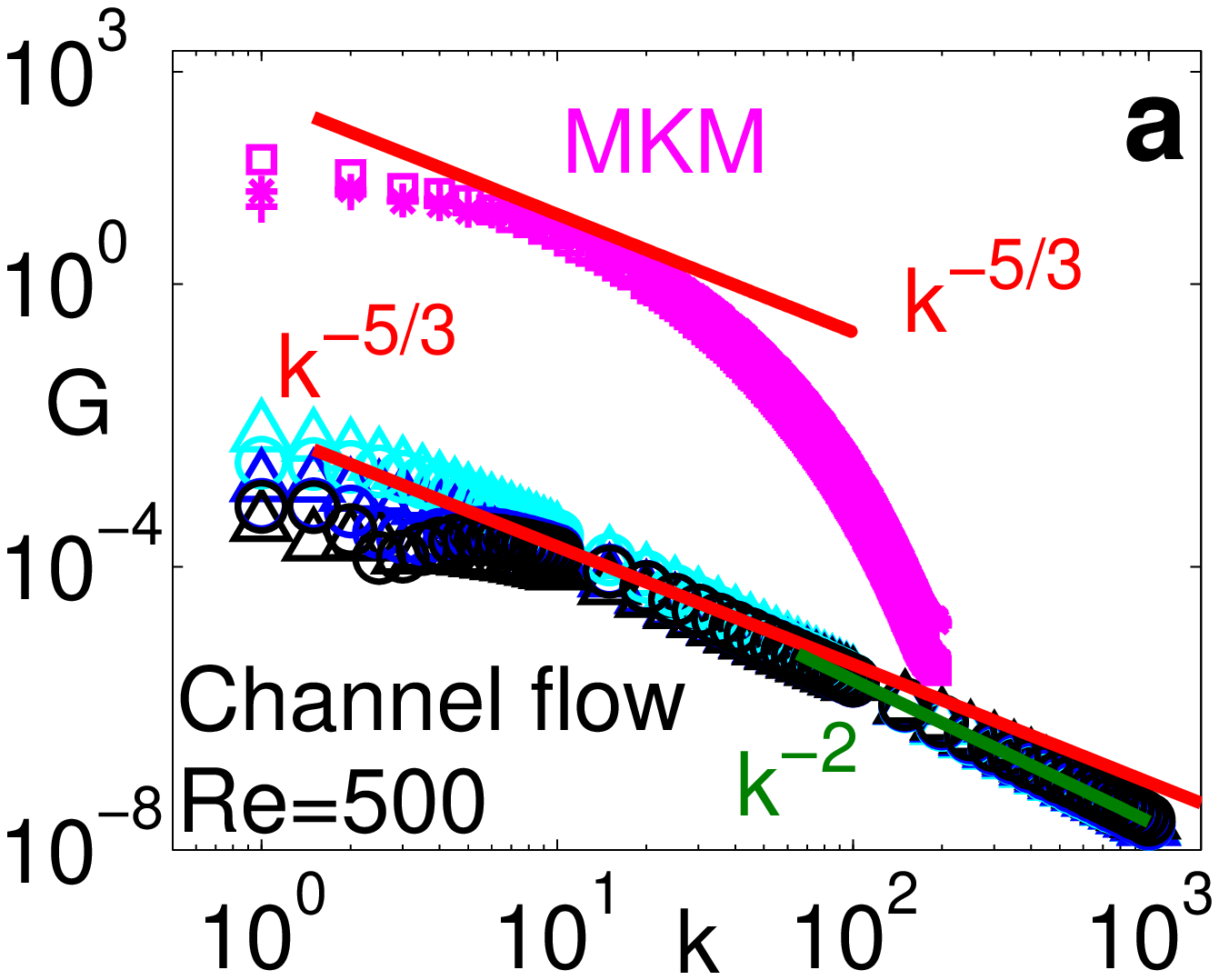}
\end{minipage}
\begin{minipage}[]{0.48\columnwidth}
	\includegraphics[width=1\columnwidth]{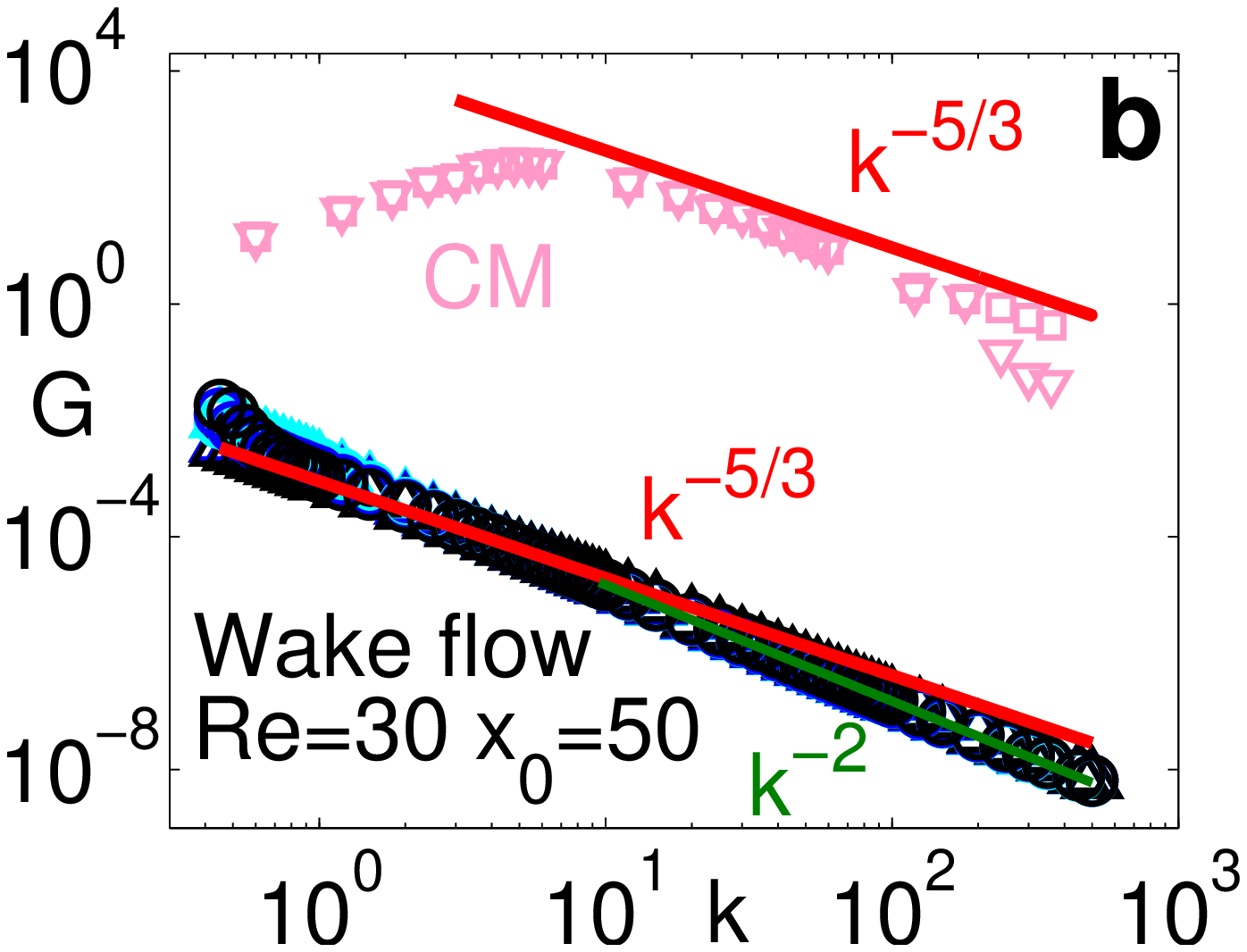}
\end{minipage}
\centering
\begin{minipage}[]{0.48\columnwidth}
	\includegraphics[width=1\columnwidth]{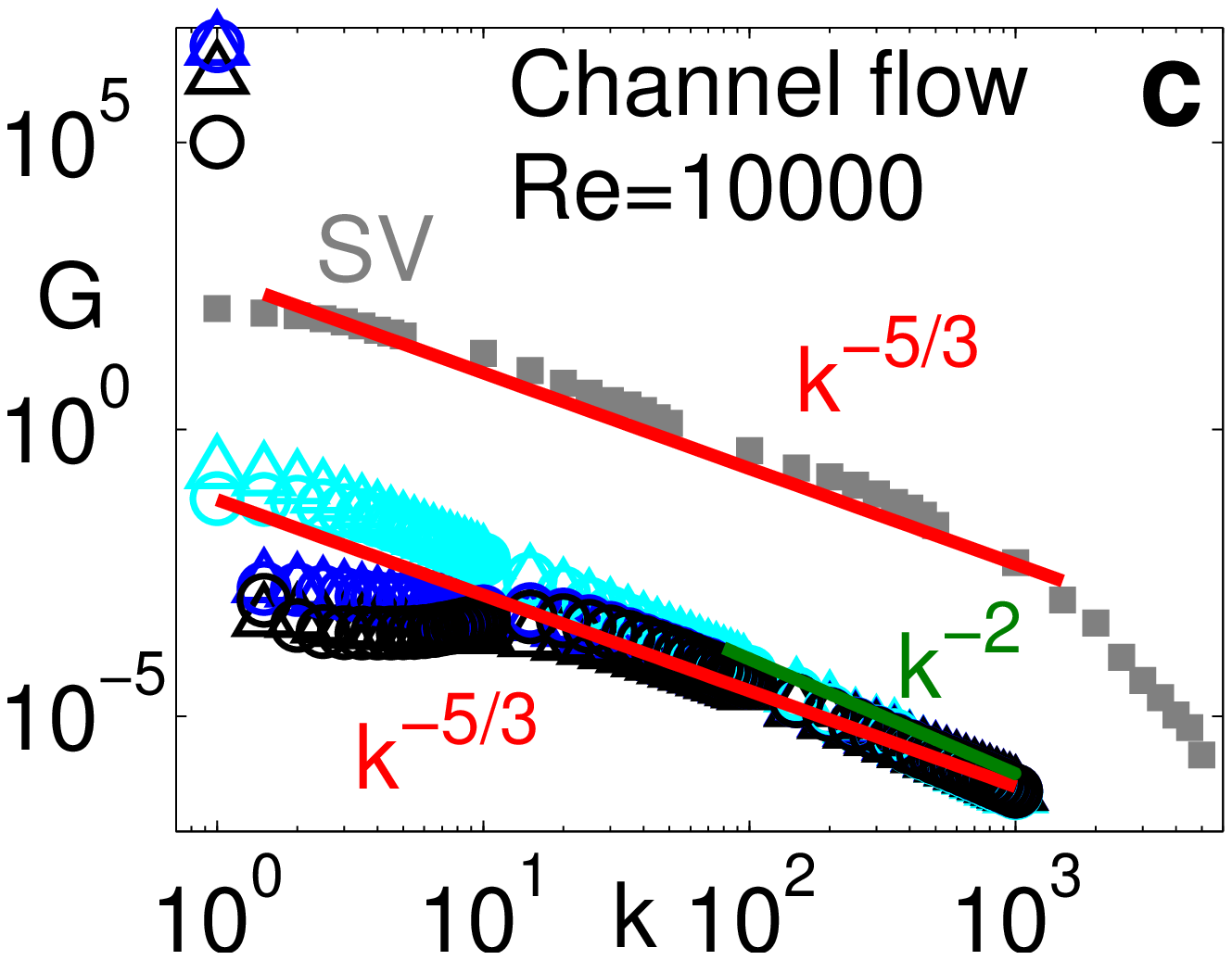}
\end{minipage}
\begin{minipage}[]{0.48\columnwidth}
	\includegraphics[width=1\columnwidth]{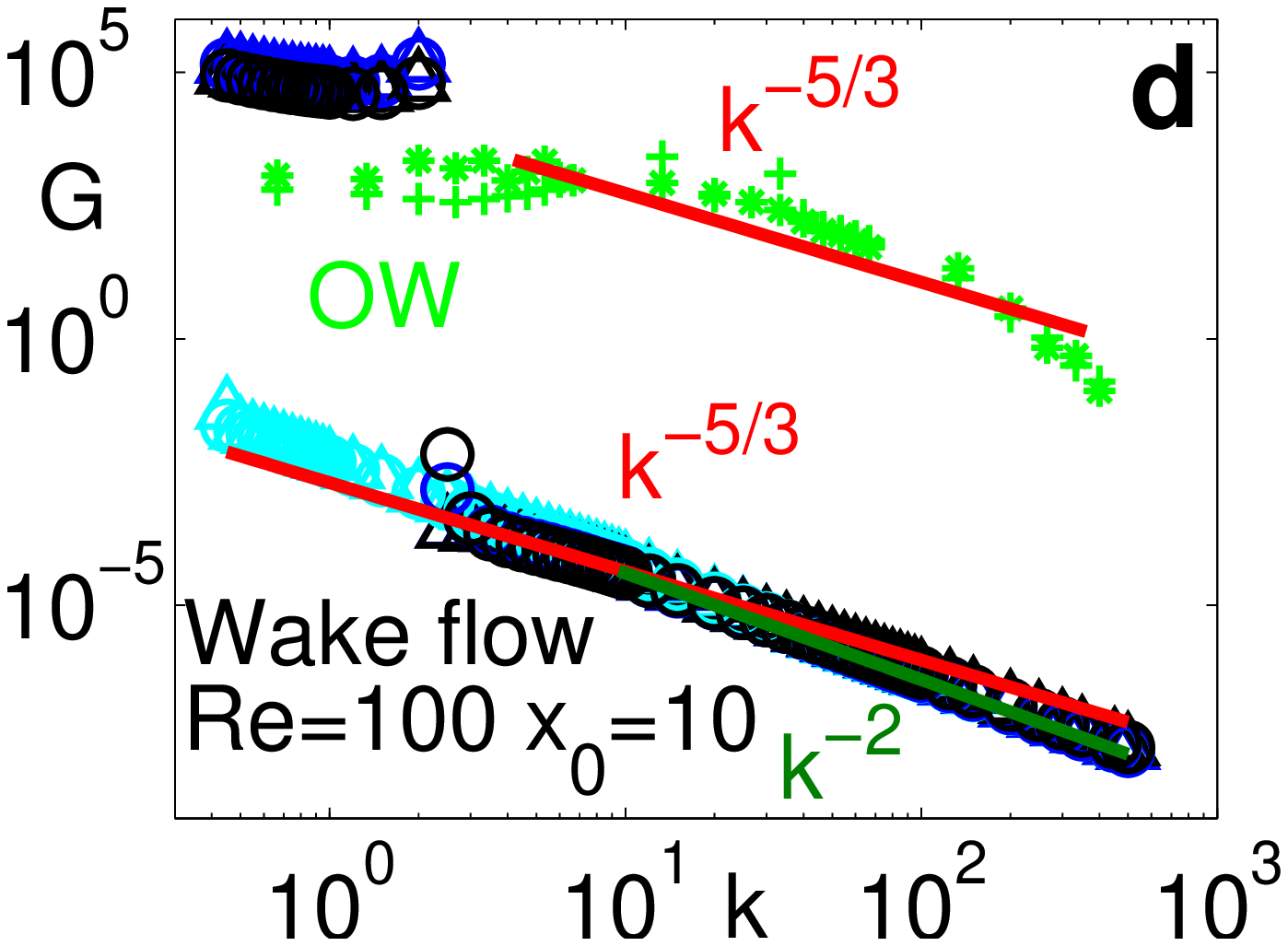}
\end{minipage}
\centering
\begin{minipage}[]{\columnwidth}
	\includegraphics[width=\columnwidth]{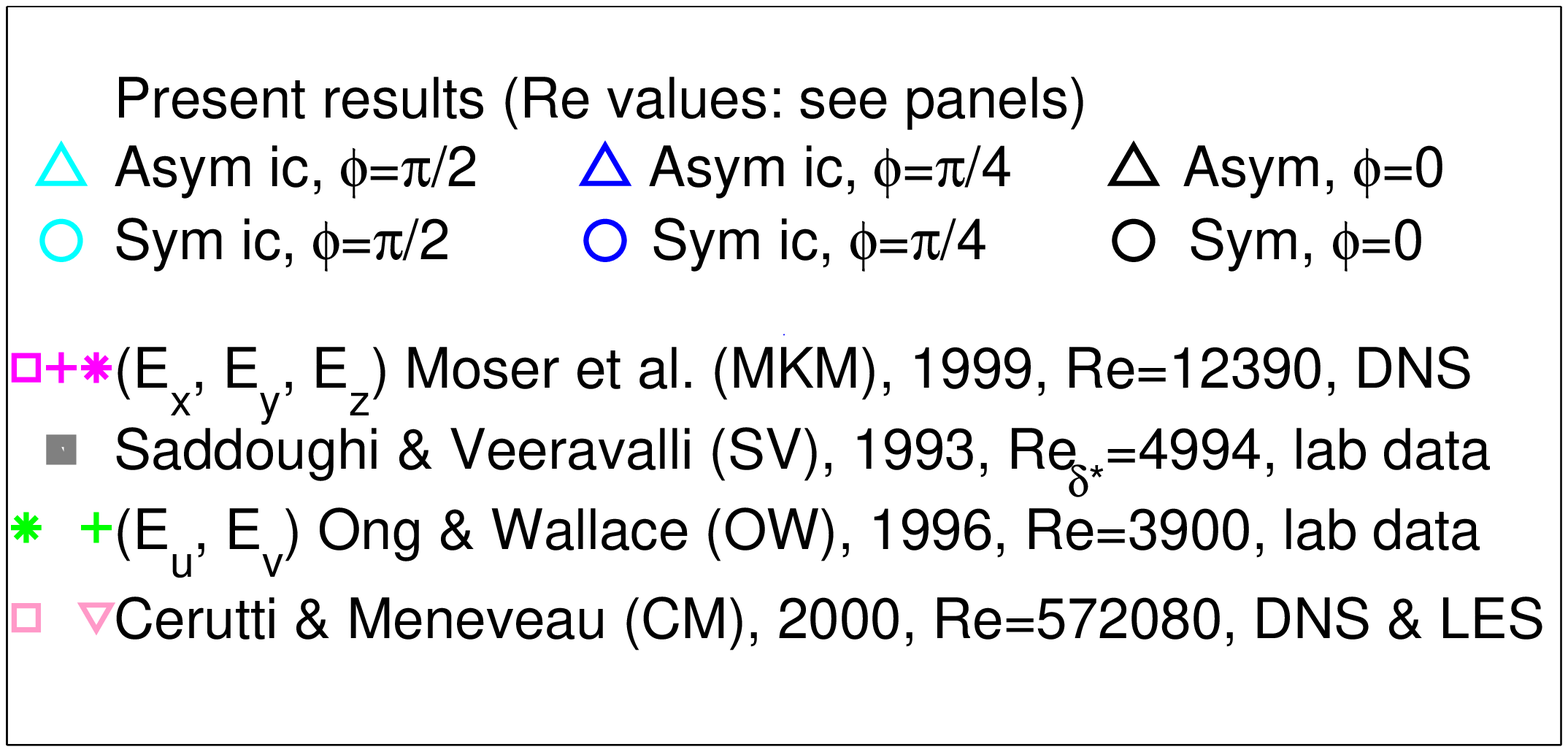}
\end{minipage}
\caption{Spectra of the amplification factor G for the collections of linear traveling waves observed at the time, $T_e$, where the perturbations  are out of their transient.
%Our results concern both stable flow configurations (panels a and b) and unstable ones (panels c and d).
The spectra obtained in this study  have been compared qualitatively with the turbulent field  spectra available in the literature and obtained from laboratory or numerical experiments.
%In this case, the spectra are obtained by means of Fourier analysis of the unsteady spatial fields frozen at given temporal instants.
The legend in the bottom panel specifies the symbols associated with either the results of the present study or those of the laboratory, and numerical experiments carried out on fully turbulent flow fields by other authors.}
\label{fig:spettri}
\end{figure}

\noindent In order to determine the temporal region in which the  evolution behaves exponentially, it is necessary to monitor the instants beyond which the condition $r \rightarrow const$ is satisfied. Doing this, we introduce a necessary condition, that, however, is not sufficient to determine the instants where the perturbations can be compared. In fact, the transition toward the exponential behavior is smooth, very long, and different from case to case (in some instances, it can be oscillatory). As a result, it is not numerically efficient to obtain the instants where  $r$ starts to be  constant.  We have thus associated a second condition to be satisfied together with the constancy of the growth rate, which directly acts on the energy temporal rate of variation. To this aim, we have selected the instants at which the amplification factor reaches a given rate of variation, either in growth  or in decay. This situation is  represented by the instant, that we call observation time, $T_e$, where $ dG/dt < \epsilon $ or $dG/dt > 1/ \epsilon$, with $\epsilon = 10^{-n}$. Here, $n$ is an arbitrary  positive quantity (for instance, a positive integer) that we have fixed equal to 4. It is possible to show that the present results -- in particular, the existence of an intermediate spectral range where the spectral decay exponent is very close to that of the Kolmogorov theory -- do not depend on the choice of $n$ (see SM, section 1).

In Fig. 3 and 4, the results of the measure of the energy residuals out of the transients  are shown. The spectral values of $G$, for both the channel flow case (panels a,c) and the wake flow case (panels b,d)  show  a scaling in the intermediate range of the polar wavenumber ($k\in[10, 200]$ for the channel, $k\in[2,50]$ for the wake) that is amazingly close to the turbulent canonical value of $-5/3$. For shorter wavelengths, characterized by very short transients, the scaling is a little higher in magnitude, approximatively equal to $-2$. This result does not appear to be influenced to any great extent by the wave obliquity, the symmetry, or  the $Re$. However, it is possible to observe that  purely orthogonal waves show a closer scaling  to $-2$ than to $-5/3$, even at intermediate wavenumbers.
In general, a full decade of intermediate wavenumbers can be observed for both the wall flow and the free flow. %, in which longitudinal and oblique perturbations present a power-law decay which is close to -5/3 (red curves), while purely orthogonal waves ($\phi = \pi/2$) have a decay of about -2 (green curves).  All the perturbations for larger wavenumbers show a power-law decay  close to -2 (green curves).
These data gather all the stable waves occurring  in the intermediate range %(which is usually named inertial in  turbulence jargon)
and in the dissipative range. We would like to point out  that the data do not highlight a dependence on $Re$, the  flow control parameter. For longer waves ($k<10$ and $k<1-2$ for the plane Poiseuille flow and the bluff-body wake, respectively), the results depend on the perturbation inclination, the symmetry of the initial condition, and on the boundary conditions (geometry of the system). As expected, they do not reveal any universal behavior. In Fig. 3, panels a-b-c-d,  experimental (laboratory and numerical) measurements \cite{MKM1999,SV1993,CM2000,OW1996} in the turbulent states have been included for the sake of comparison of stable linear perturbations and turbulent scales.

%It should be noted that
These results appear strengthened by the fact that  even the observation times, $T_e$, present the same scaling, see Fig. 4.  This outcome is not at all trivial. It is sufficient to consider that the observation time includes the transient, and that the different kind of transients we observed are very complex and can vary in length over 4 orders of magnitude when moving across the space of parameters (the wavenumber, the symmetry, the obliquity, the flow control parameter, Re, and, in the case of a basic evolving spatial flow, the position).

\begin{figure}
\centering
\begin{minipage}[]{0.48\columnwidth}
	\includegraphics[width=1\columnwidth]{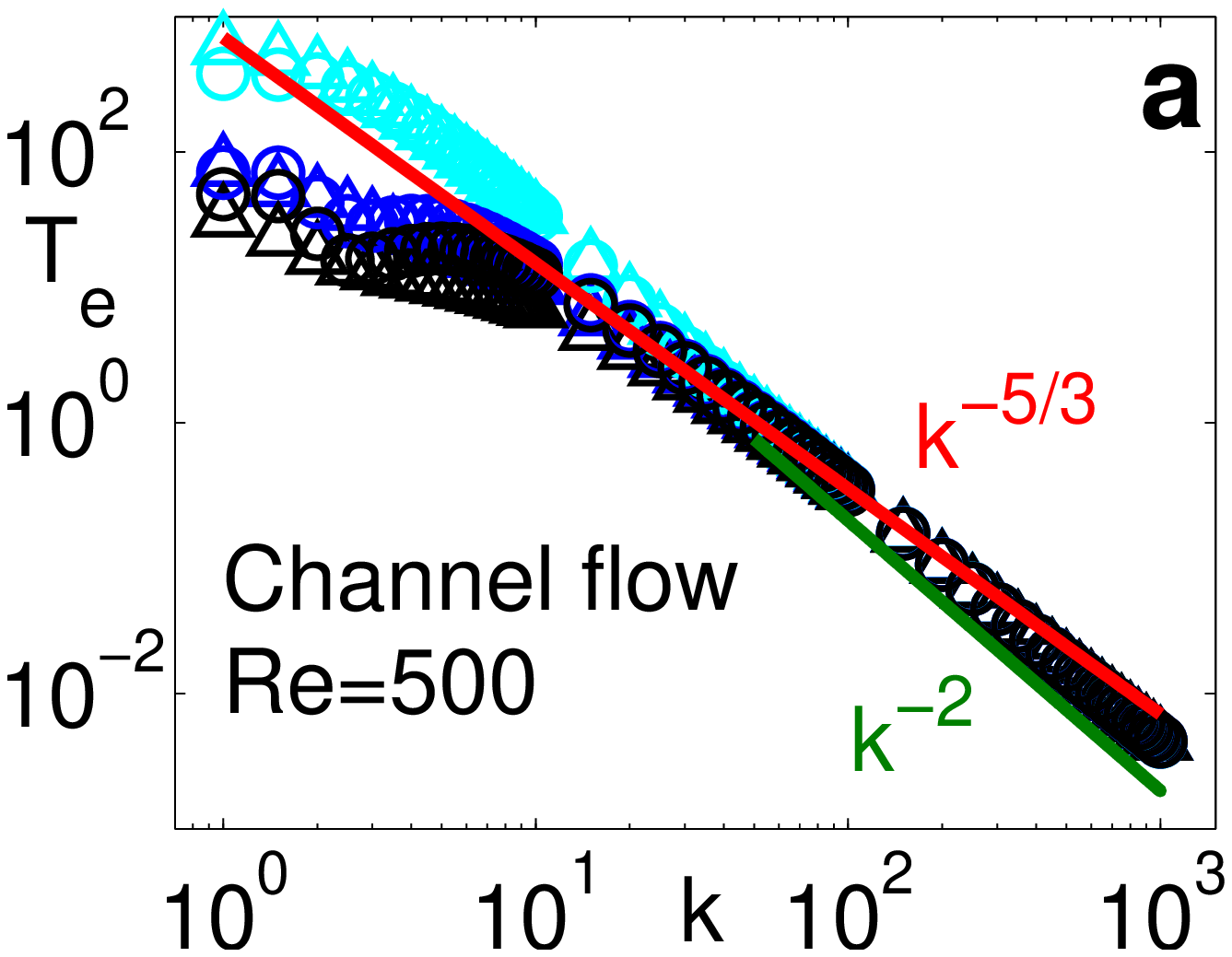}
\end{minipage}
\begin{minipage}[]{0.48\columnwidth}
	\includegraphics[width=1\columnwidth]{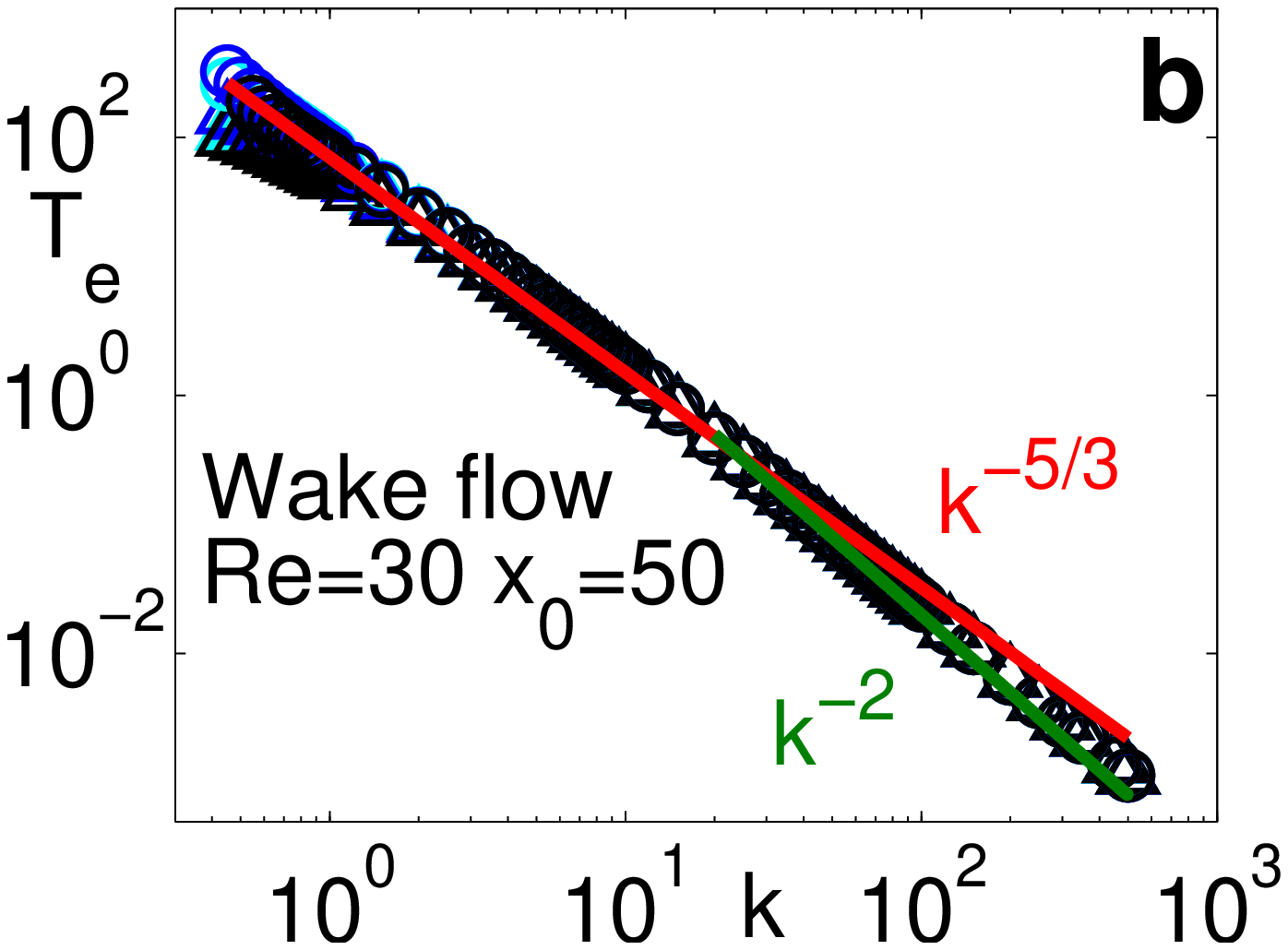}
\end{minipage}
\begin{minipage}[]{0.48\columnwidth}
	\includegraphics[width=1\columnwidth]{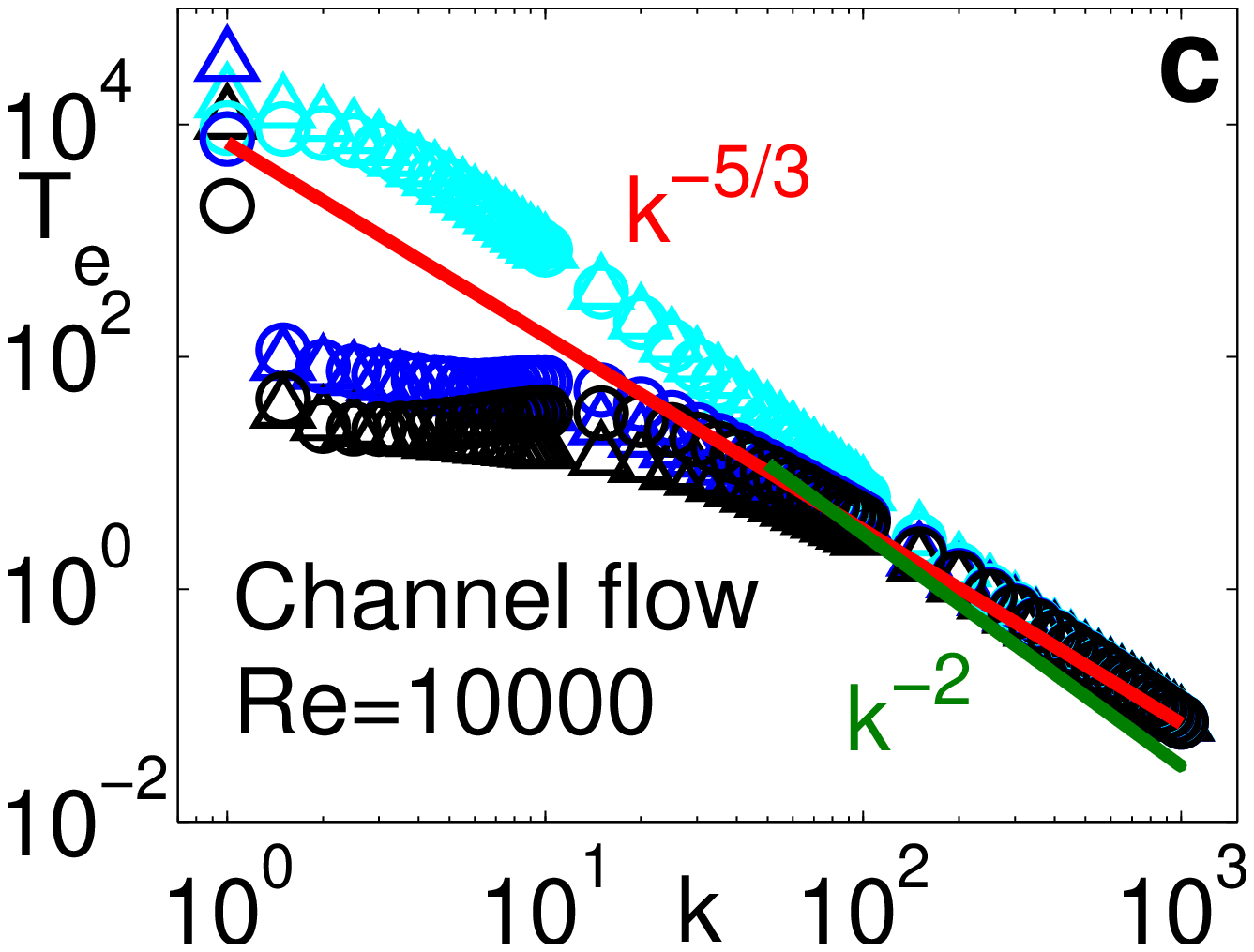}
\end{minipage}
\begin{minipage}[]{0.48\columnwidth}
	\includegraphics[width=1\columnwidth]{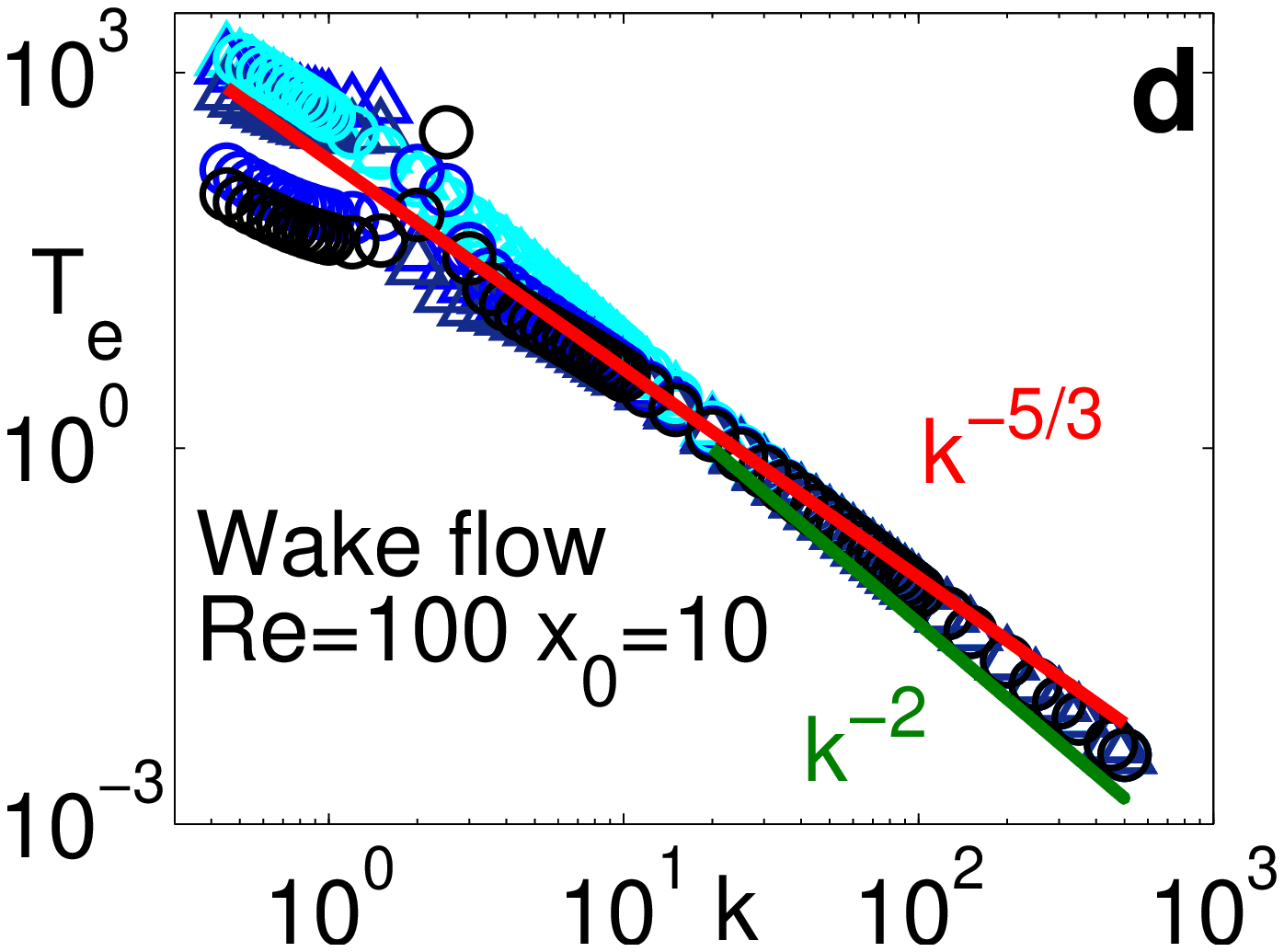}
\end{minipage}
\centering
\begin{minipage}[]{\columnwidth}
	\includegraphics[width=\columnwidth]{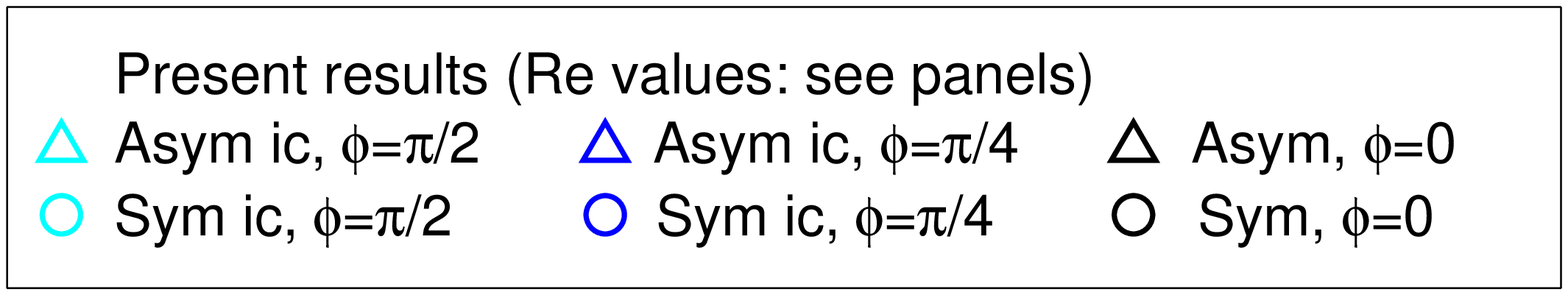}
\end{minipage}
\caption{Spectrum of the observation time where the wave energy is measured and compared. The data refer to the entire collection of linear perturbations shown in figure 3.  Stable basic flows (top panels),  unstable flows (bottom panels). The plane channel flow is on the left, the bluff body wake is on the right. Symmetric (circles) and asymmetric (triangles) initial perturbations. Obliquity angles: $\phi=0$ (black), $\phi=\pi/4$ (blue) and $\phi=\pi/2$ (cyan). It is possible to observe  an abundant decade where longitudinal, oblique and orthogonal perturbations present a power-law decay which is close to -5/3 (red curves). Larger wavenumbers, which are  influenced more by diffusion-dissipative processes, steep towards a decay exponent of about -2.}
\label{fig:Te}
\end{figure}

A possible scenario for fields with a sufficiently high Reynolds number to be in a turbulent state,  which includes the present findings, could be the following. The nonlinear interaction works intensively on the long waves that are unstable by blocking the linear  growth and by settling their kinetic energy to a level that can be associated with the global Reynolds number of the system. The surplus of energy that the unstable waves potentially gain during the linear evolution is then transferred to shorter waves, that are  longitudinal, oblique or orthogonal to the basic velocity field. The transfer is particularly intense for  asymptotically stable intermediate waves (namely the inertial scales, according to the terminology used in turbulence). It is possible to suppose that the transfer is physically triggered  during the  phases where the amplification factor is maximum, see Figure 2. %As a consequence, the nonlinear interaction allows the raising of the energy of the intermediate waves to a level comparable to that of the longest ones, see in figure 3 the curves representing the turbulent spectra. %It is more probable that the nonlinear coupling becomes active in these phases.
This process continues and the energy in the range of intermediate  waves reaches  values that can be experimentally observed,  in the laboratory, or in the so called Navier-Stokes direct numerical solutions.
Taking into account the present results, it is possible to  say that the nonlinear interaction distributes the relative energy over different wavenumbers in a way that corresponds to the relative residual energy each wave has  when, after the transient, it reaches a common decay threshold.
This process is less efficient on short waves. For these waves, the maximum of the amplification factor becomes progressively smaller as the wavenumber magnitude increases (see Figure 2b)  and, as a consequence, it is more difficult for them to couple with longer waves.
A rapid fall in energy to levels many orders of magnitude lower, and, thus  closer to the specific levels  of their linear evolution, is therefore observed in the experimental spectra (see Figure 3).

In conclusion, we have observed a fingerprint of the Kolmogorov scaling inside the collective behavior of transient intermediate  perturbation waves, which always are asymptotically stable. These new observations are not specific of a peculiar kind of flow (wall bounded or free). This can  mean that this scaling is not only one of the major signatures of the turbulence interaction, but it also exists  hidden inside the dynamics of linear stable waves, where even the self-interaction is absent.  %One can then conjecture how many other properties commonly believed specific of turbulence could be found  in the linear  perturbation dynamics.
%Once known, these properties could induce an important outcome: the modeling and forecast of turbulence fields could be done on the basis of linear solutions quickly numerically obtained  and in general more  physically  inclusive  than contemporary turbulence models. The use of super-computers could become  less strategic in the future.  Linear solutions in fact do not need  parallel computations, and can often  be preliminary analytically treated in a way where at least one or two  spatial variables can be  removed from the differential system of governing equations.
Since our observations do not depend on the system control parameter (Reynolds number), on the kind of initial condition and on geometrical parameters, such as the wave inclination, they could  also reveal a new set of structural properties of the Navier-Stokes equation solutions. In particular, we think  that  they can be used to build a bridge between the linear and  the nonlinear interaction in multi-scale systems.
Given the high variability of the shape and length of the transient life, we consider remarkable that the Kolmogorov scaling does not appear only in the energy spectra, but also in the spectra of the observation times.  %All the more, regardless the possible values taken by the geometrical and dynamical parameters. At large, we think that our findings can suggest that the spectral energy scaling commonly observed in fully turbulent flows is a more general dynamical feature that links  the linear evolution of perturbation waves, which are stable in the longterm, to inertial nonlinear coupling.

%This sheds some light on the existence of some common linear and nonlinear Navier-Stokes solutions properties, and helps one to consider them as a whole. In particular, we consider it is important to explore the collective behaviour of the linear solutions by thinking of them as a set of multiple elements that fill the wavenumber range which pertains to the geometry of the system. By means of a comparison of the various elements, or, possibly,  by means of an observation of their collective behaviour, this multiplicity may either reveal trends that are common to the nonlinear dynamics or highlight their differences.

For the critical and useful exchanges had during her visit, D.T. acknowledges the 2011 program The Nature of Turbulence, held at the Kavli Institute for Theoretical Physics of the University of California Santa Barbara. For helpful comments D.T. acknowledges Katepalli R. Sreenivasan, and, together with S.S., William O. Criminale and B. Eckhardt. The authors thank Marco Mastinu for his contribution to the numerical simulations. This study is in part supported by the Progetto Lagrange Foundation.

\end{document}

\vspace*{0.5cm}
\begin{center}
{\huge Supplemental Material}\\[1cm]
{\huge\bfseries {Kolmogorov scaling bridges linear hydrodynamic stability and turbulence}}\\[2cm]
{\Large Stefania Scarsoglio, Francesca De Santi, Daniela Tordella}\\[0.5cm]
{\large email: \ttfamily daniela.tordella@polito.it}
\end{center}

\tableofcontents
%\vspace*{1.5cm}
%car\noindent
%\section*{This PDF file includes the following material: DA FARE ALLA FINE}
%\begin{itemize}
%\item Construction of the mixing layer: initial conditions and transient (Figures S.1 and S.2, Table S.1).
%\item Why a Reynolds stress and a mean flow are not generated inside the shearless mixing layer? (Figure S.3).
%\item Transversal velocity derivative skewness (Figure S.4).
%\item Figures 1, 2, 3
%\item Table 1

%\end{itemize}

\vspace*{0.5cm}

\section{Wave comparison criterium. Residual energy out of the transient.}

Figure S1, see in particular the caption content, shows that the  double criterium used to define the observation time $T_e$, where we  measure  the energy of the traveling waves, is not conditioning the existence and the exponent of the spectral scaling presented in Figs. 3 and 4 of the main paper, and in Fig. S2 of the following section.

\begin{figure}
\centering
\begin{minipage}[]{0.48\columnwidth}
	\includegraphics[width=\columnwidth]{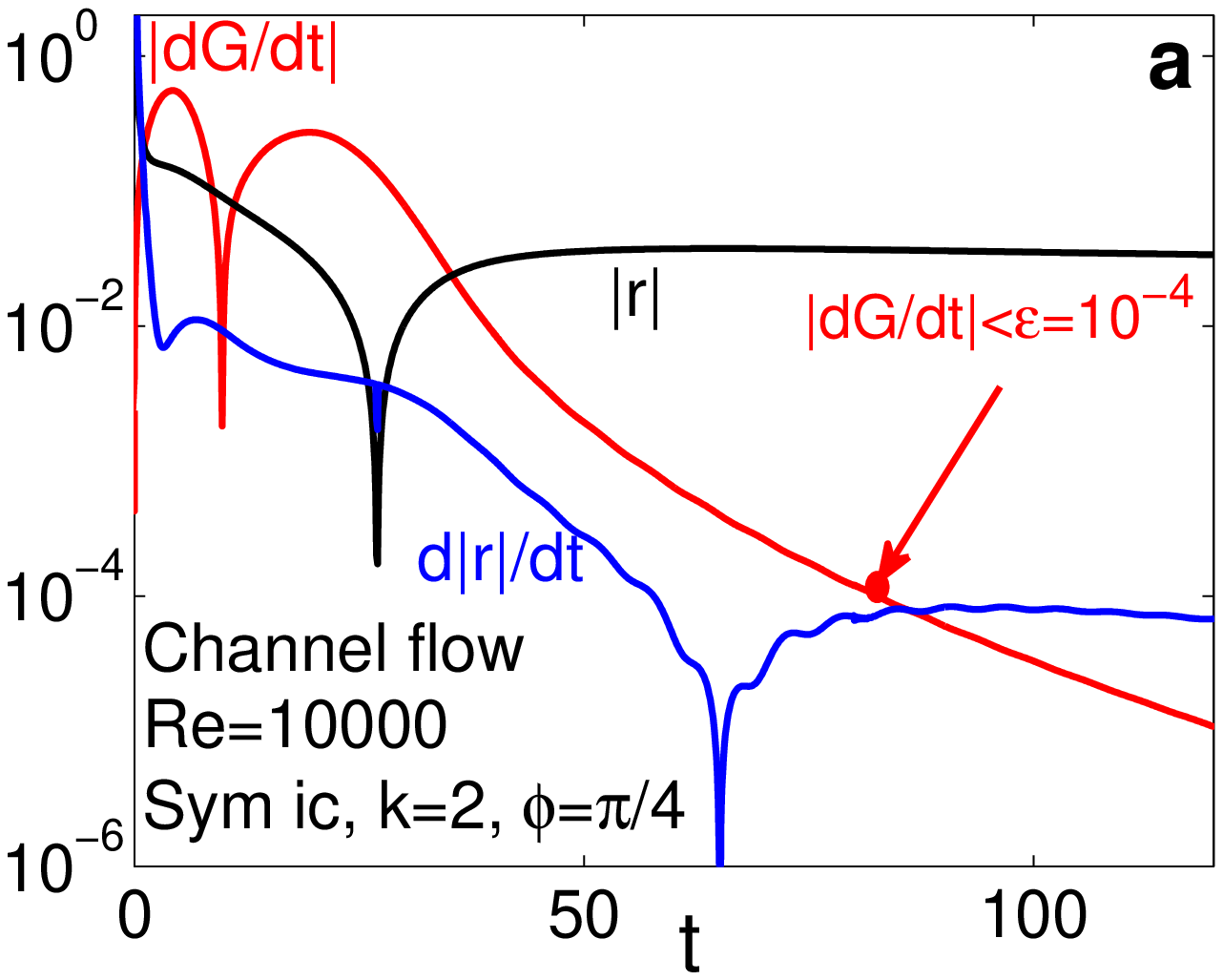}
\end{minipage}
\begin{minipage}[]{0.48\columnwidth}
	\includegraphics[width=\columnwidth]{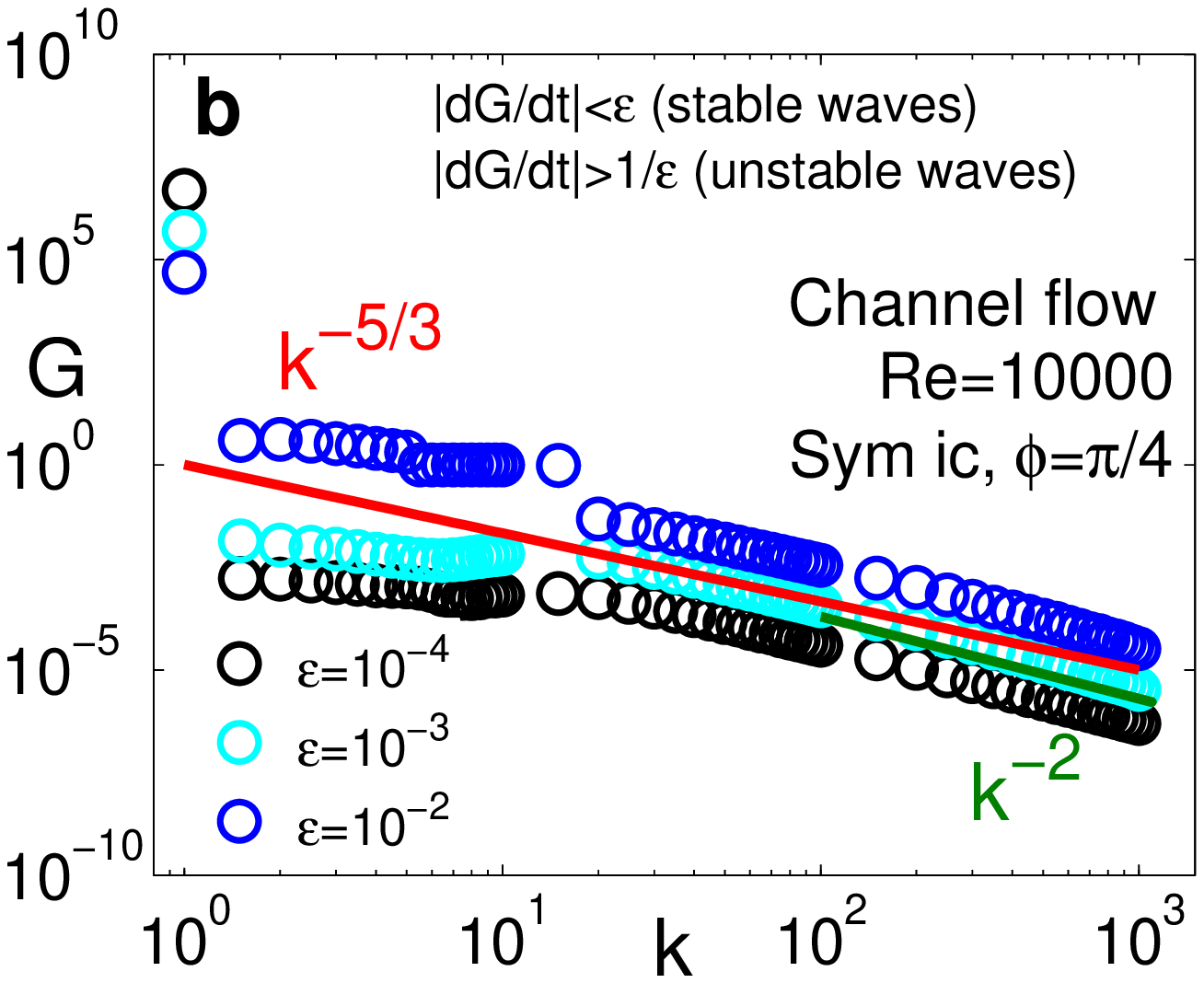}
\end{minipage}
\centering
\begin{minipage}[]{0.48\columnwidth}
	\includegraphics[width=\columnwidth]{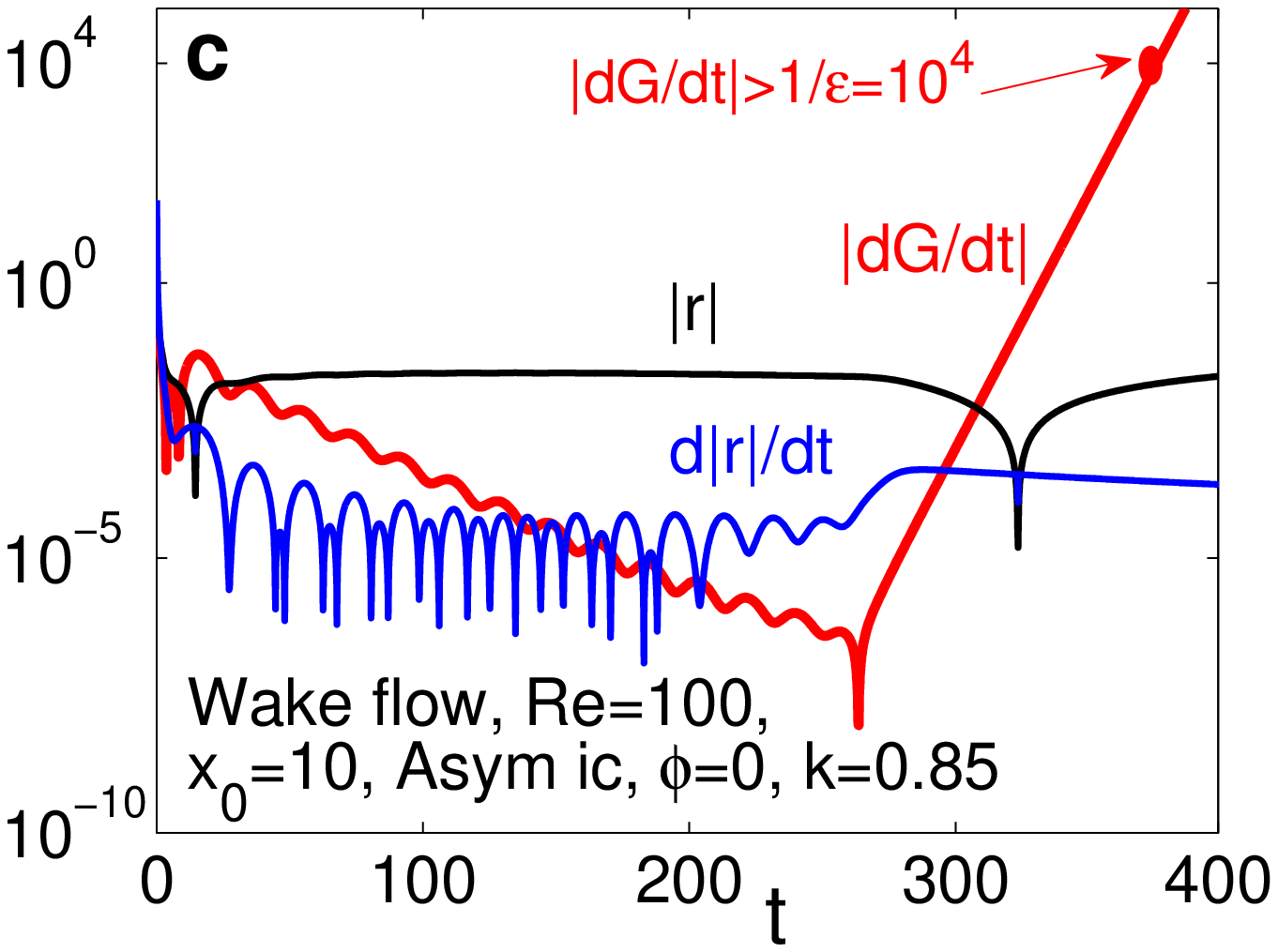}
\end{minipage}
\begin{minipage}[]{0.48\columnwidth}
	\includegraphics[width=\columnwidth]{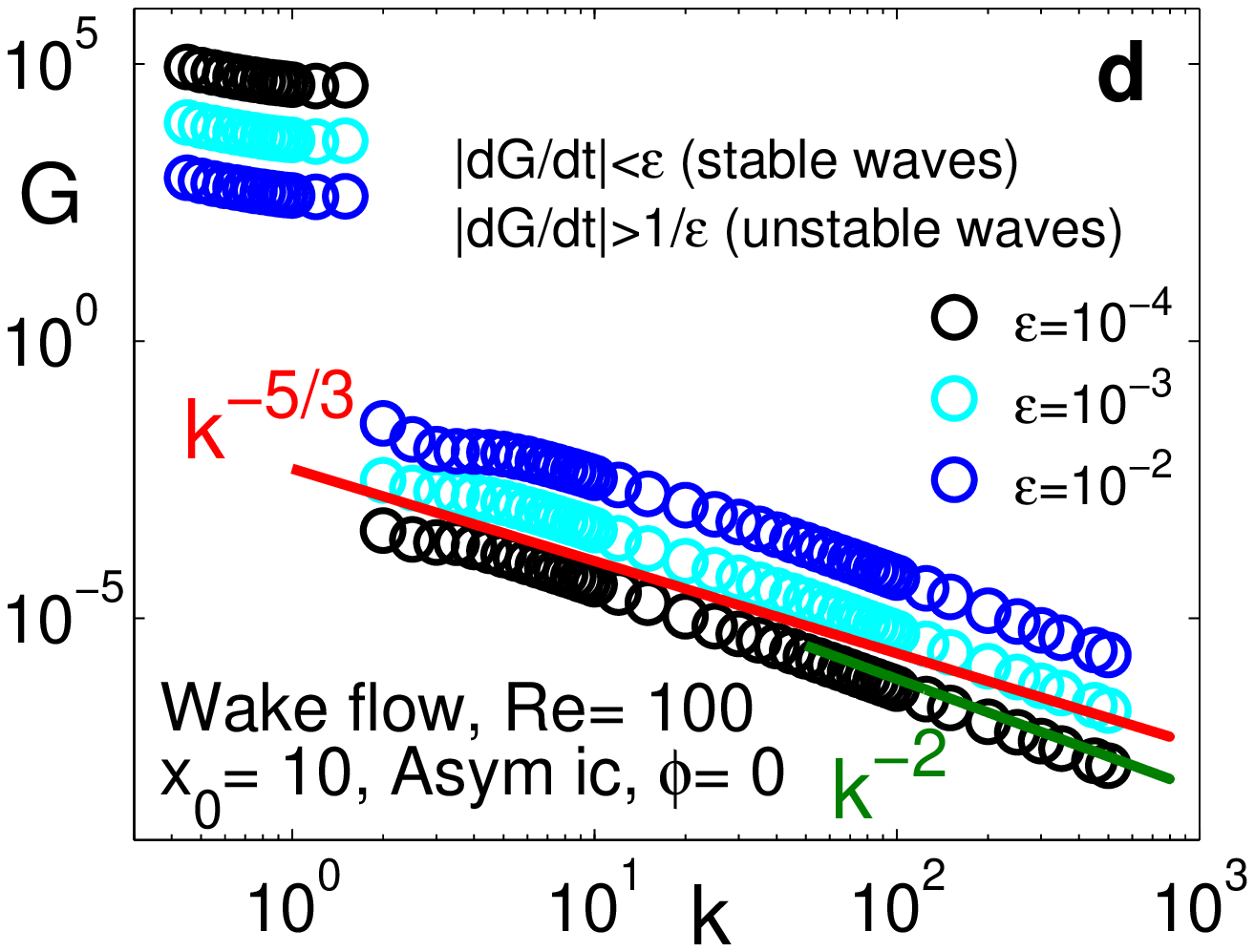}
\end{minipage}
\label{fig:as}
\caption{Criteria to define the exit time from the transient part of the perturbation life. (a)-(c) channel and wake flows:  we consider  here $G$, the amplification factor, i.e the energy normalized over the value taken at the initial instant,  and $r$, the temporal growth rate $
r(t; \alpha, \gamma) = log(e)/(2t)$, and show their time dependence:  -- (black curves) $|r|$, -- (blue curves) $d|r|/dt$, -- (red curves) $|dG/dt|$).
In order to determine the temporal region in which the  evolution behaves exponentially, it is necessary to monitor the instants in which the condition $r \rightarrow const$ is satisfied. Since this transition is smooth, we need the help of a further condition to  precisely define an instant at which the wave energy can be compared. To this aim, we have selected the instants at which the amplification factor reaches a given rate of variation either in growth  or in decay. This situation can be represented by the instant, that we call observation time, $T_e$, where $ dG/dt < \epsilon $ or $dG/dt > 1/ \epsilon$, with $\epsilon = 10^{-n}$, where $n$ is an arbitrary  positive quantity (for instance, a positive integer). Panels (b) and (d) in this figure show that the present results do not depend on the choice of $n$. In this example, we have tested the values $n=2, 3$ and $4$.}
\end{figure}

\section{Additional spectral distributions}

\begin{figure}
\centering
\begin{minipage}[]{0.48\columnwidth}
	\includegraphics[width=\columnwidth]{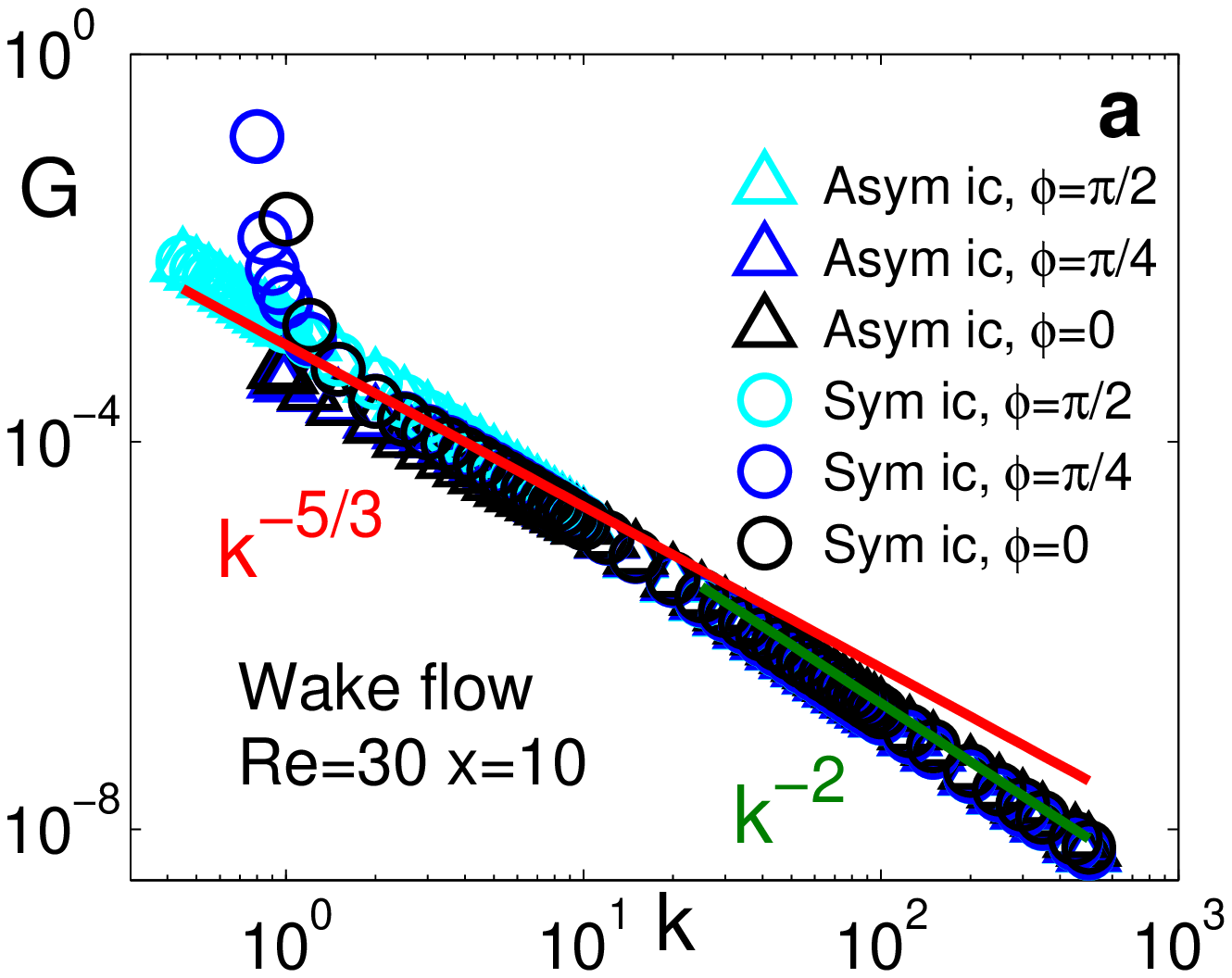}
\end{minipage}
\begin{minipage}[]{0.48\columnwidth}
	\includegraphics[width=\columnwidth]{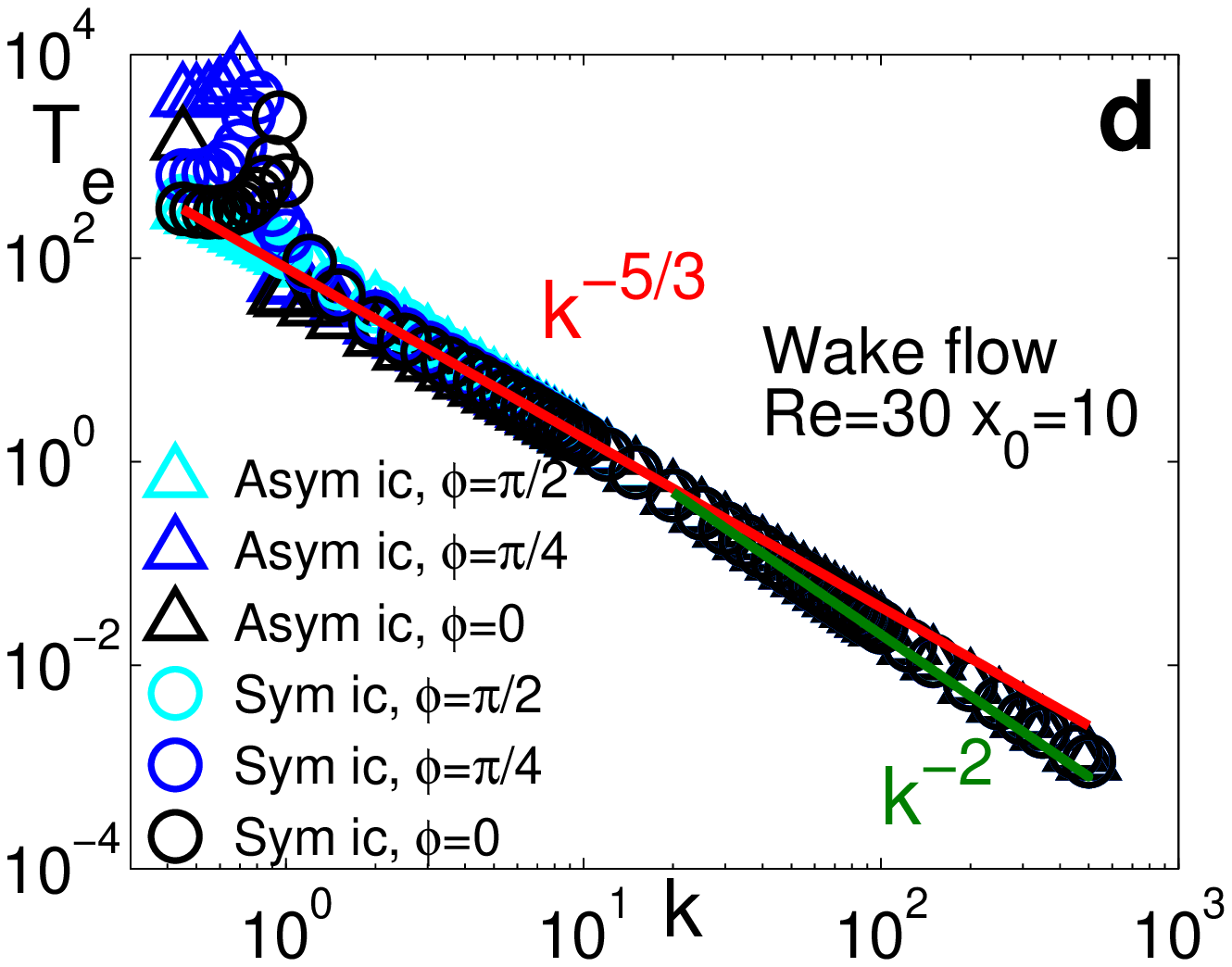}
\end{minipage}
\centering
\begin{minipage}[]{0.48\columnwidth}
	\includegraphics[width=\columnwidth]{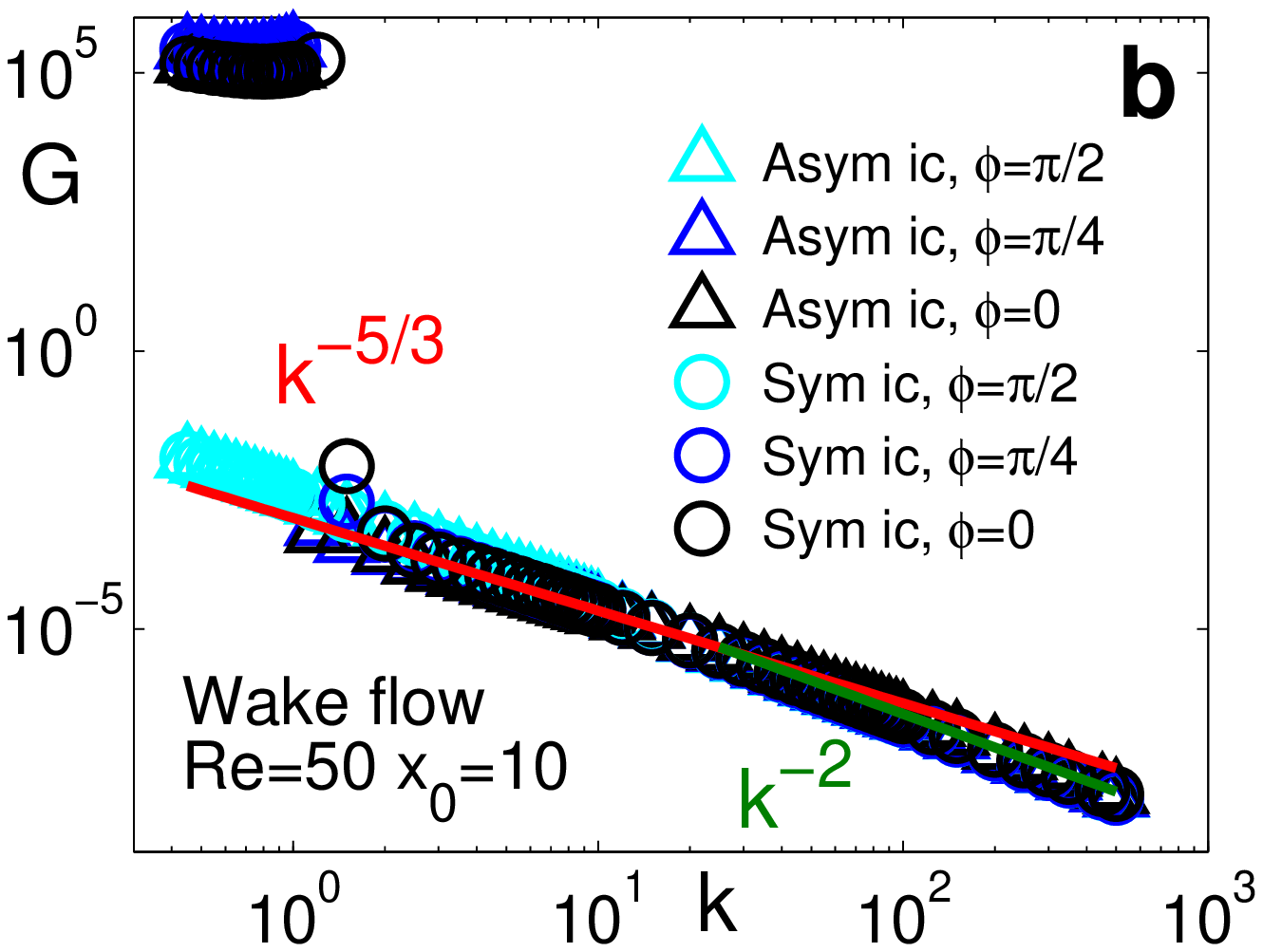}
\end{minipage}
\begin{minipage}[]{0.48\columnwidth}
	\includegraphics[width=\columnwidth]{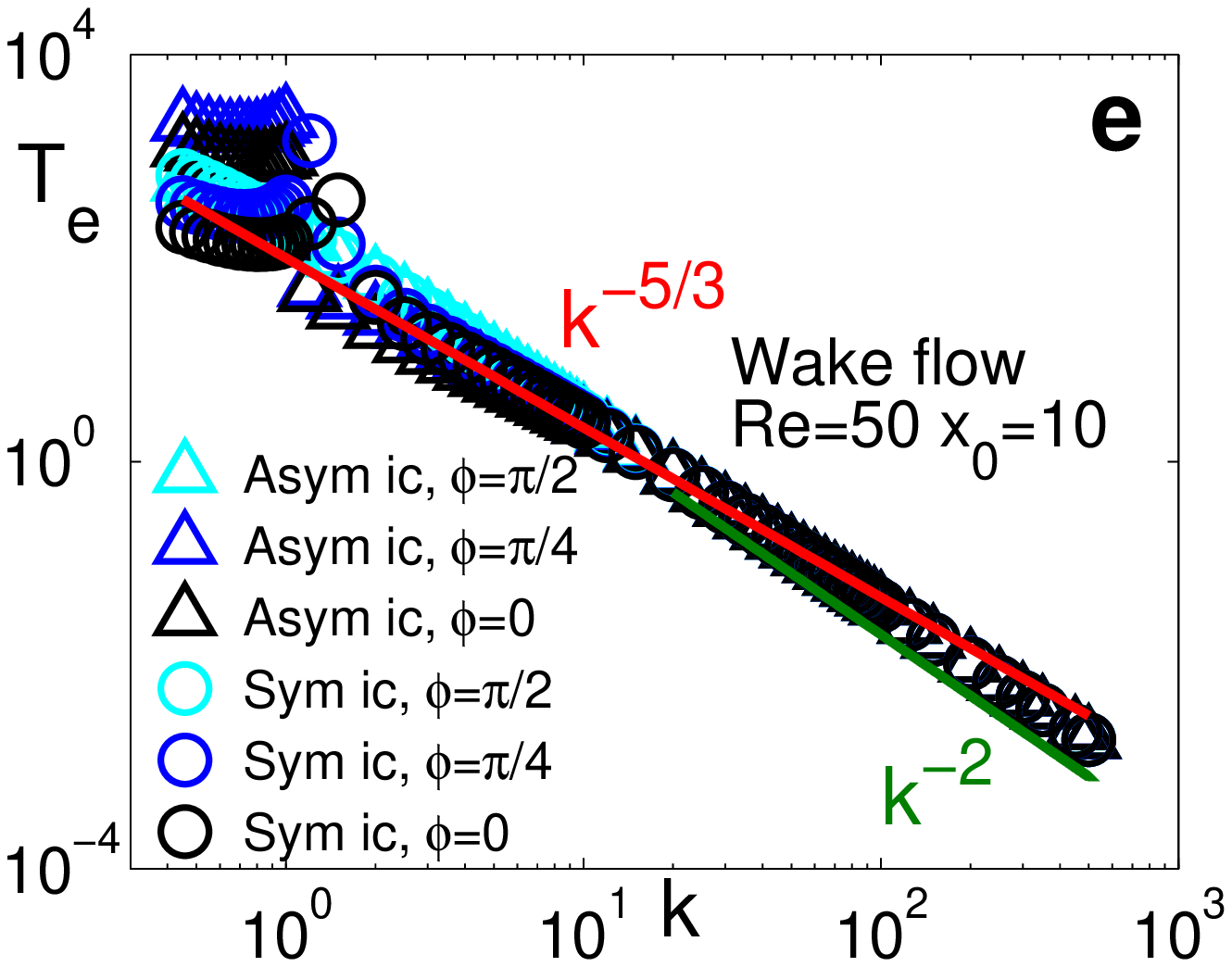}
\end{minipage}
\centering
\begin{minipage}[]{0.48\columnwidth}
	\includegraphics[width=\columnwidth]{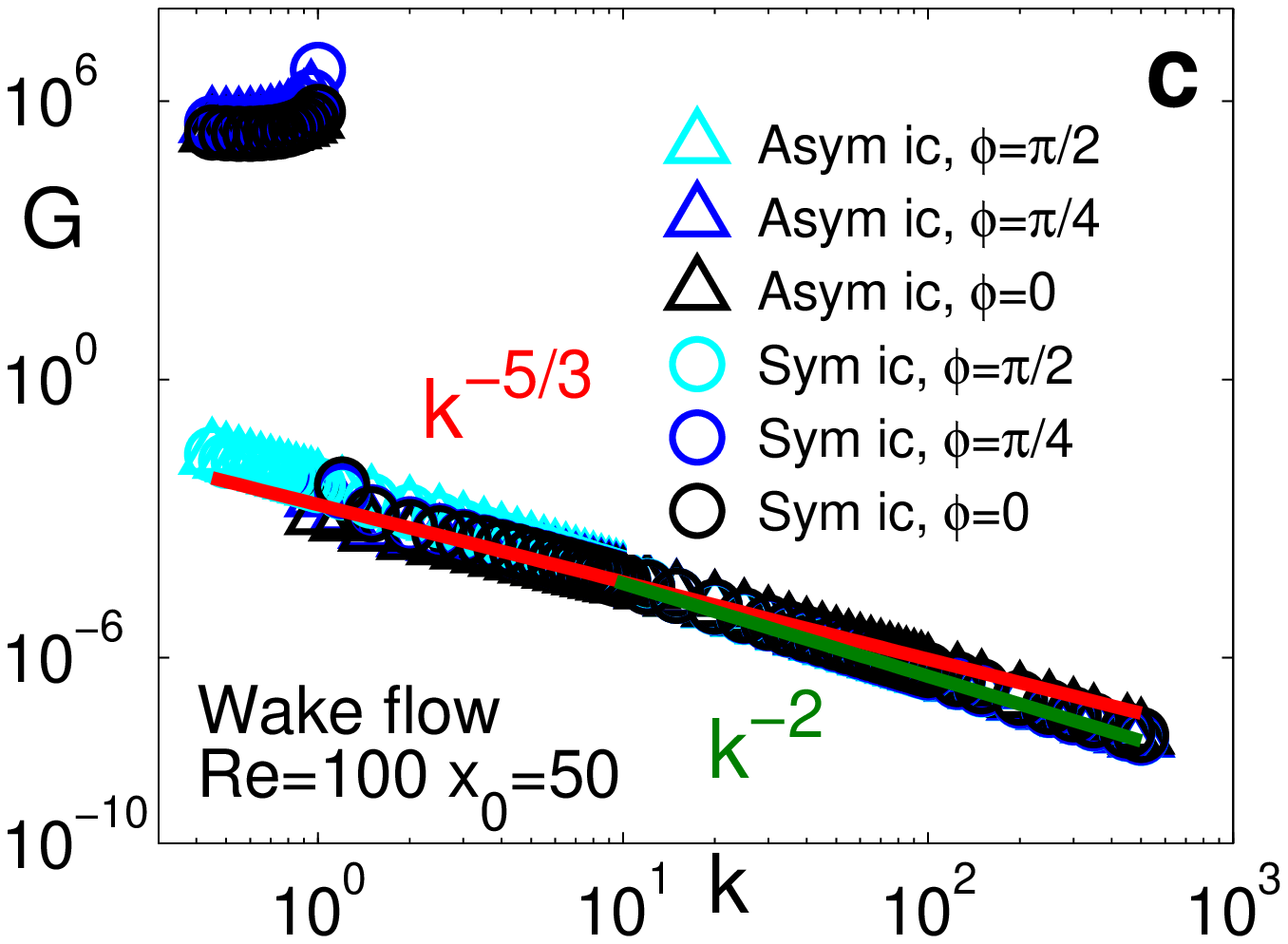}
\end{minipage}
\begin{minipage}[]{0.48\columnwidth}
	\includegraphics[width=\columnwidth]{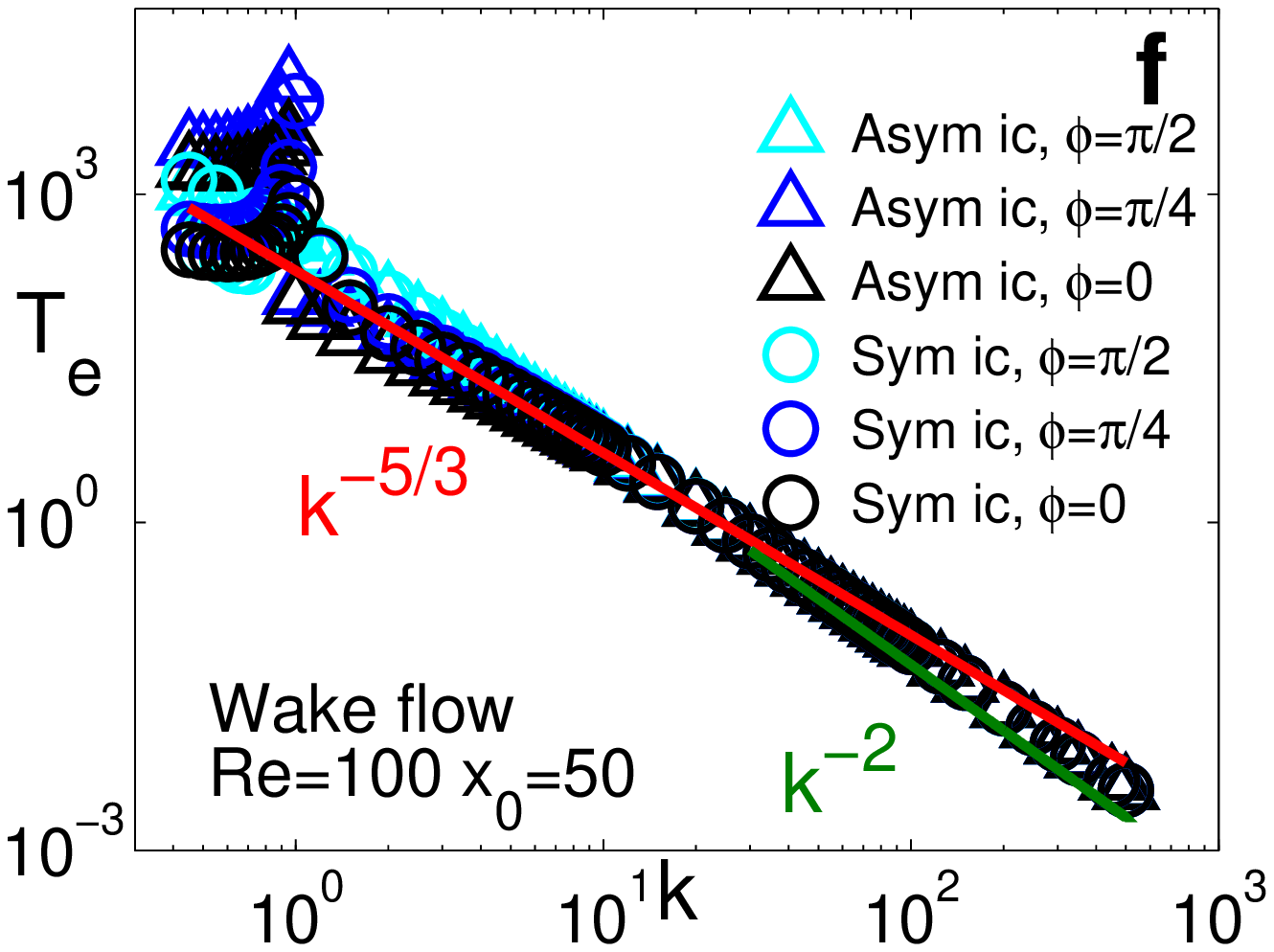}
	\end{minipage}
\label{fig:spettri}
\caption{Additional spectral distributions for the wake flow. (left) Spectra of the amplification factors $G$ and (right) spectra of $T_e$. (\textbf{a})-(\textbf{d}) $Re=30$, $x_0=10$. (\textbf{b})-(\textbf{e}) $Re=50$, $x_0=10$. (\textbf{c})-(\textbf{f}) $Re=100$, $x_0=50$.}
\end{figure}

In  figure S2, the reader can find further data on both the energy and observation time spectra in the wake.  With regard to Fig. 3,  in the main text, one can see how the energy, in particular that of the  unstable waves, is growing with $Re$ and  decreasing with the distance from the body. %In panel A, $Re=30, x_0=10$ is subcritical: the %yellow region highlights the range of unstable wavenumbers which are not observed in the laboratory, and, as a consequence, should be discarded. It should be %noted that subcritical instability is often predicted by linear stability studies using either local or global methods, see for instance %\cite{Pier2002,HM1990,GL2007}. This is an outcome due to the poor approximation of the basic flow close to the body, which is highly influenced by entrainment %\cite{TS09}, and thus by the lateral  field dynamics. To overcome this problem the basic flow should retain the information on the transversal dynamics, as %represented by the lateral velocity. In other words the assumption of parallelism for the basic field should be relaxed. This has been demonstrated by means of %multiscale modal analyses where the contribution of the transversal velocity field in the basic  flow was included \cite{TSB06,BT06}. For computational %simplicity, in this study  we have adopted the parallelism assumption, which is working very well in both the intermediate and far field for supercritical %Reynolds numbers. For the subcritical ones, it is anyway working very well in the region downstream 10-15 body lengths. However, in the region across 10 body %scales, we observe that in subcritical condition, leaving aside the longest wave numbers, this assumption provides correct stability results for waves with %$k>1$. As expected, in fact, these waves are stable.

\section{Asymptotic pulsation}

\begin{figure}
\centering
\begin{minipage}[]{0.49\columnwidth}
	\includegraphics[width=\columnwidth]{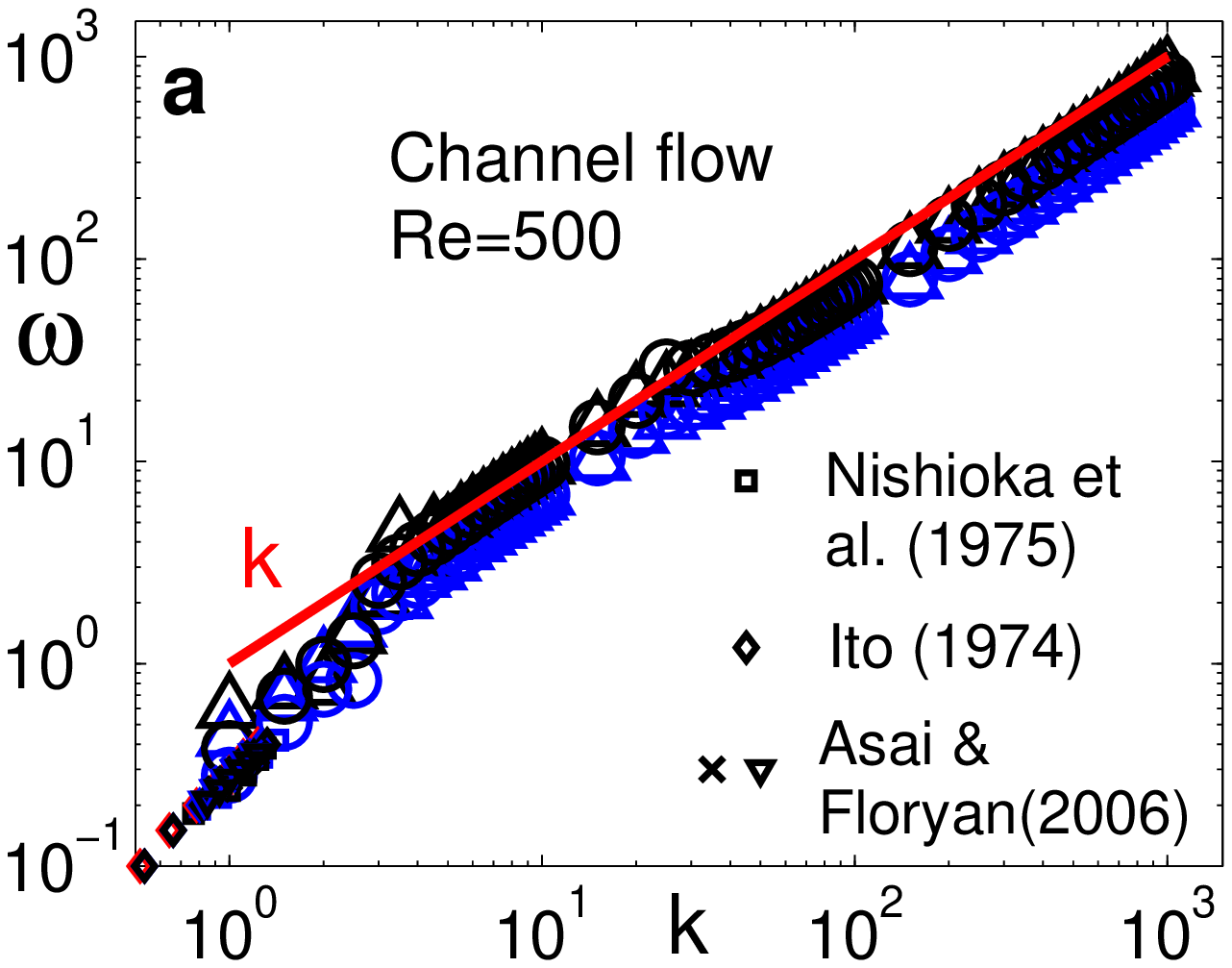}
\end{minipage}
\begin{minipage}[]{0.49\columnwidth}
	\includegraphics[width=\columnwidth]{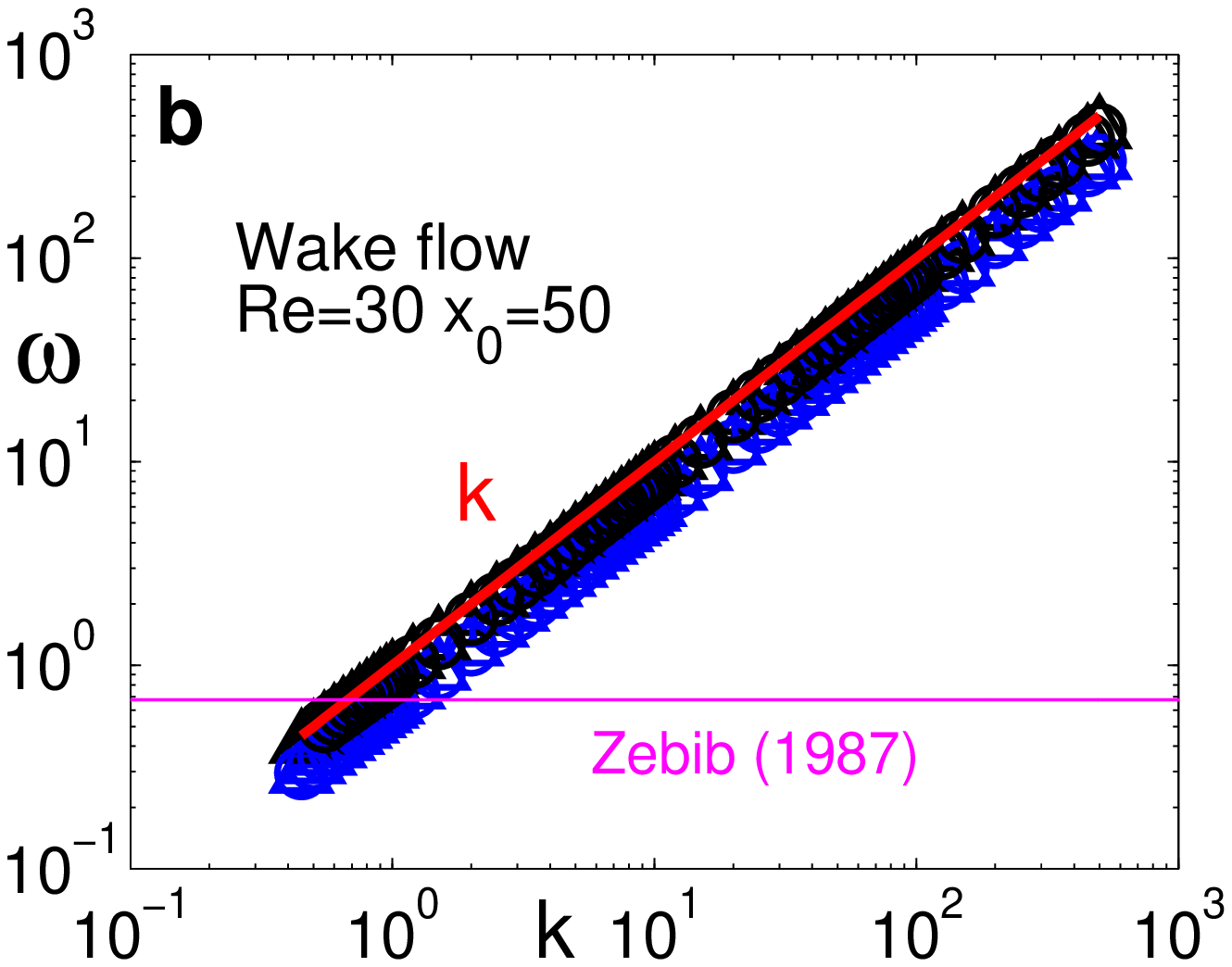}
\end{minipage}
\centering
\begin{minipage}[]{0.49\columnwidth}
	\includegraphics[width=\columnwidth]{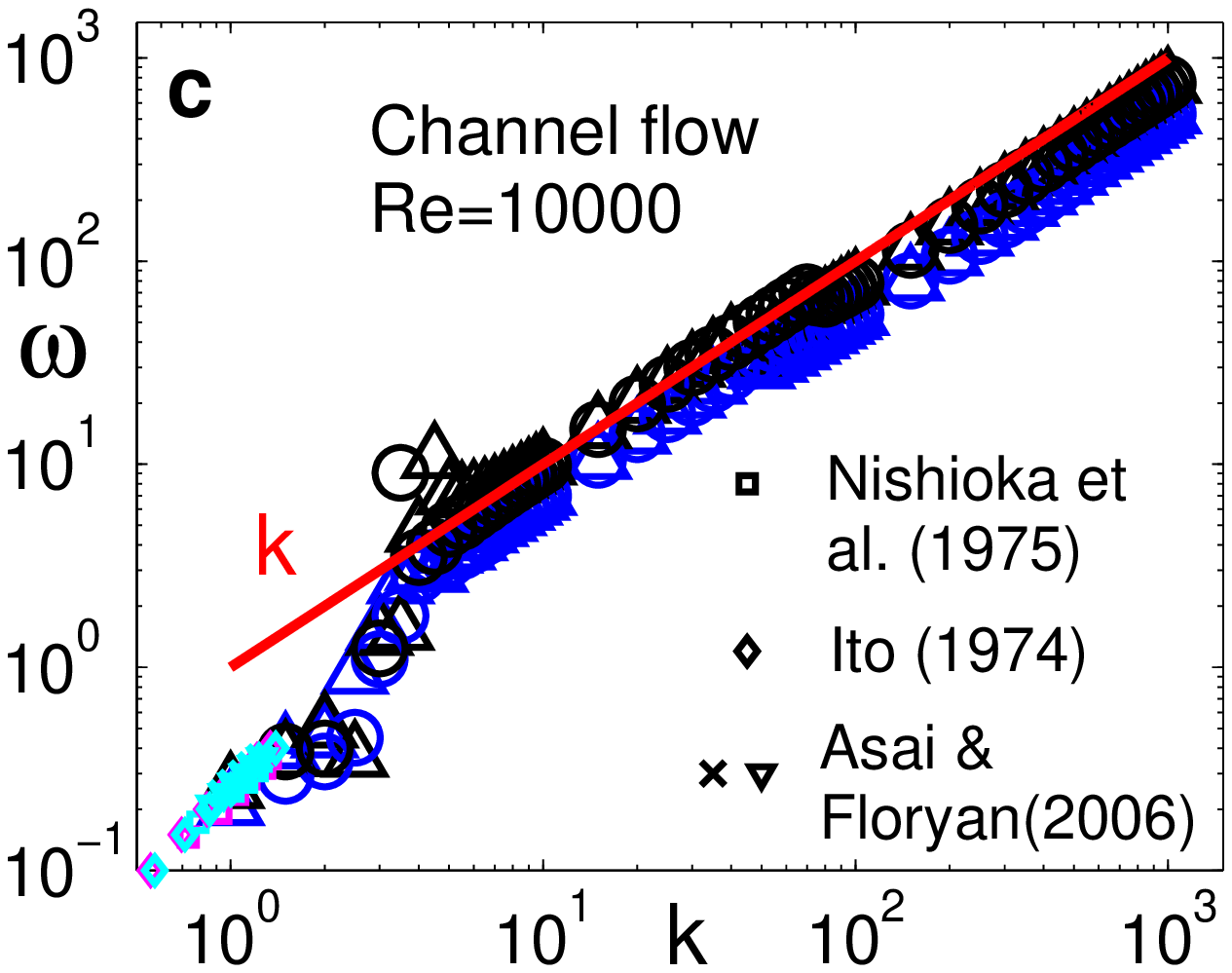}
\end{minipage}
\begin{minipage}[]{0.49\columnwidth}
	\includegraphics[width=\columnwidth]{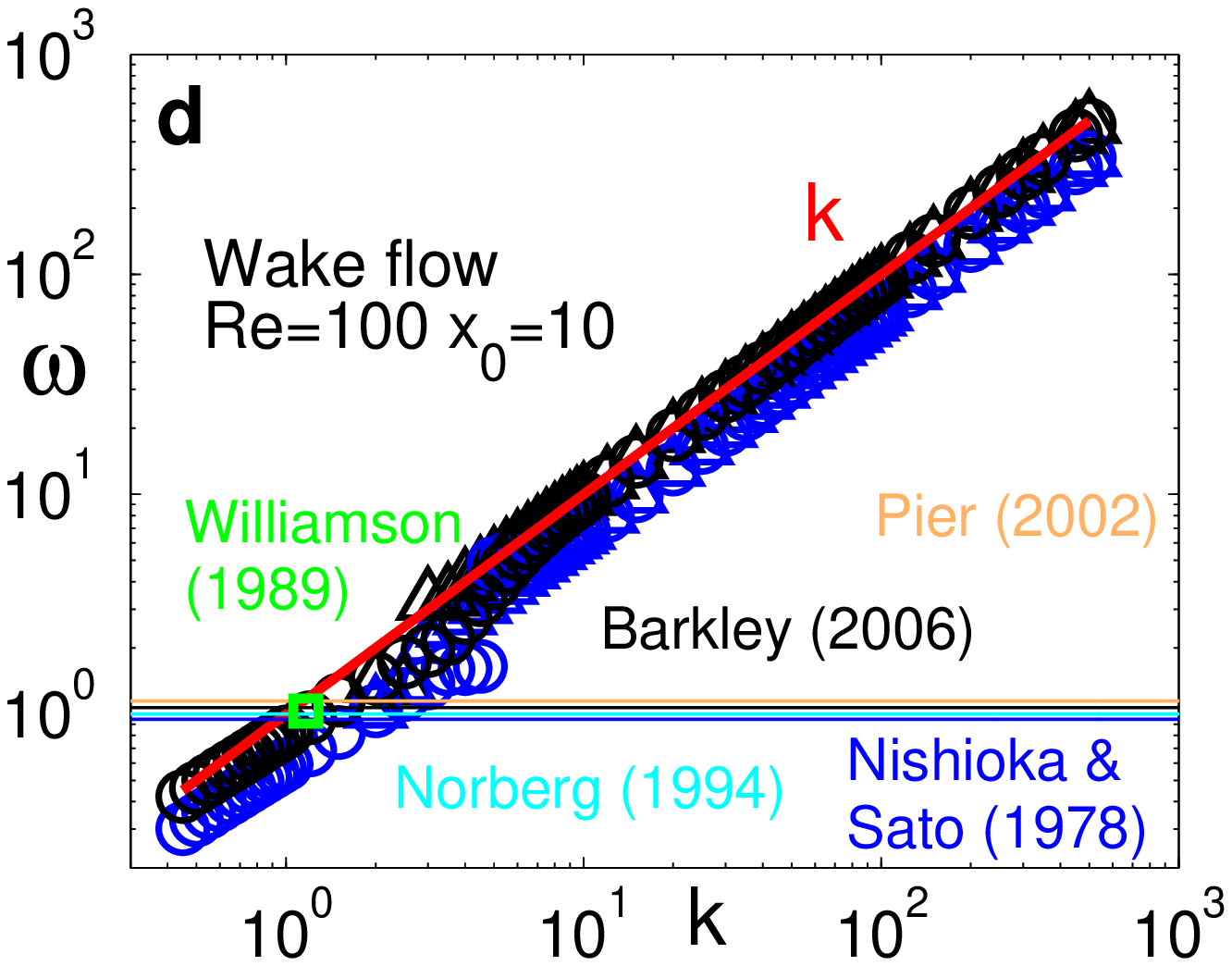}
\end{minipage}
\centering
\begin{minipage}[]{\columnwidth}
	\includegraphics[width=\columnwidth]{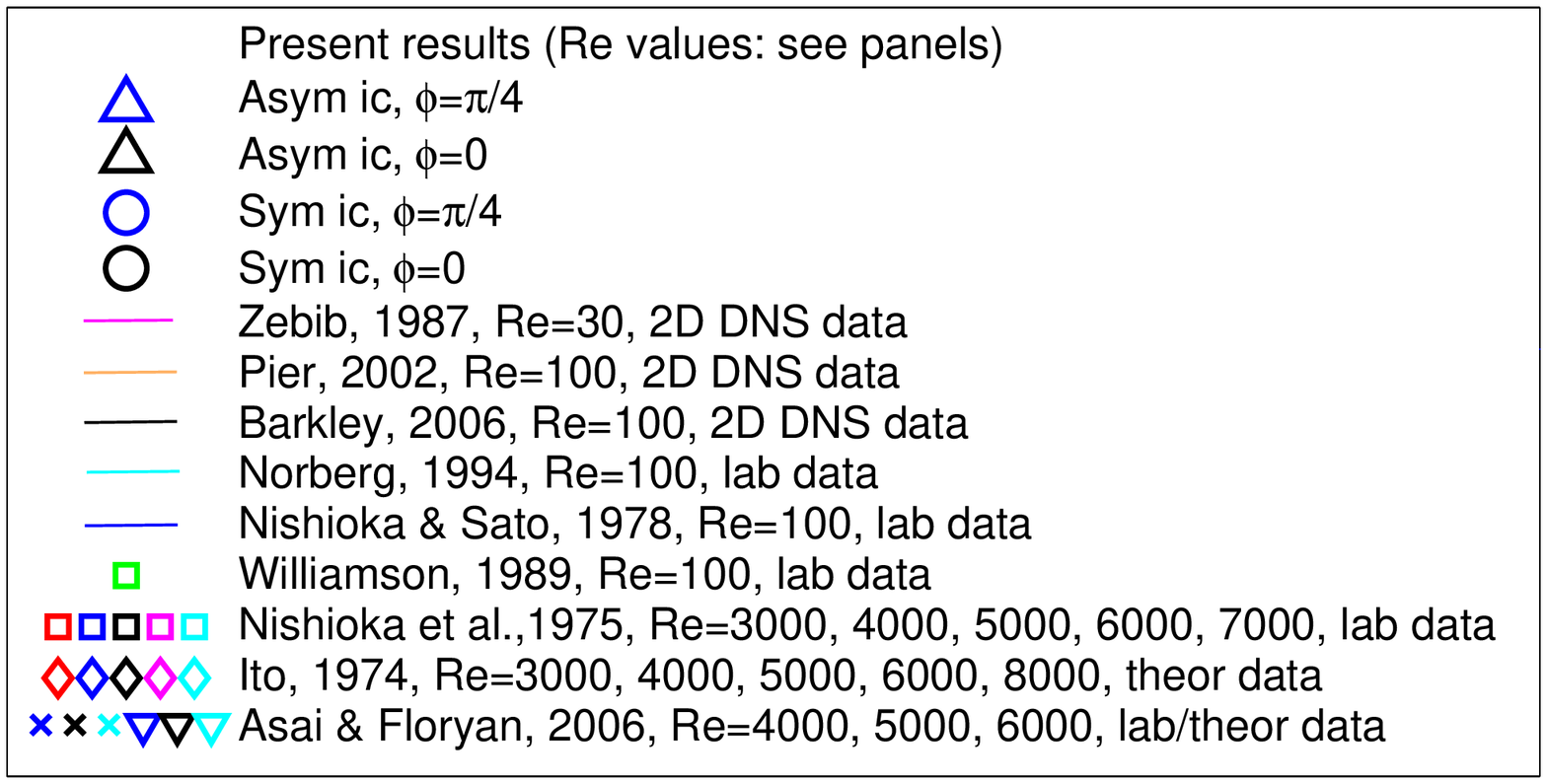}
\end{minipage}
\caption{Spectrum in the wavenumber space of the pulsation of the waves shown in figure 3 of the main text. Stable basic flows (top panels),  unstable flows (bottom panels). The plane channel flow is on the left, the bluff body wake is on the right. Symmetric (circles) and asymmetric (triangles) initial perturbations. Obliquity angles: $\phi=0$ (black) and $\phi=\pi/4$ (blue). The legend in the bottom panel specifies the symbols associated with either the results of the present study or these of the laboratory, together with numerical experiments and theoretical analysis carried out by other authors (high Reynolds number channel flows, normally in turbulent configuration, are kept laminar).}
\label{fig:omega}
\end{figure}

The spectral distribution of the pulsation is presented for the collection of linear perturbations shown in Fig. 3 of the main text. The distribution highlights a proportionality relationship, i.e.  nondispersive behavior. No experimental results covering the three decades that we can observe are available in the literature. However, some data are available for the longer wave range, the most unstable, thus the most frequently studied in the field of hydrodynamic stability %\cite{SH2001,D2002,CJJ2003},
\cite{W89,Pier2002,Bark2006,Z1987,Nor1993,NiSa1978,Ni1975,Ito1975,Asai2006}, see Fig. S3 (channel: panels a, c; wake: panels b, d). The contrast is very good. It should be noted, that only Williamson \cite{W89} presented  a complete datum  for the wake where the frequency of the wave is coupled to the value of its wavelength. For the other results, the assumption is made that the measured frequency  is globally relevant for the flow in near critical conditions, although a critical wavenumber is not specified. We can observe  two abundant decades where longitudinal and oblique waves present almost universal behavior, regardless of the symmetry of the initial condition and the Reynolds number: the frequency is proportional to the wavenumber (red curves, the exponent is $1$). It is possible to observe the same behavior for the orthogonal waves (not reported here),  but the pulsation values are 20 orders of magnitude lower (the values range at around $10^{-20}$). This means that orthogonal waves, when in asymptotic conditions - i.e. outside their transients - become standing waves.

\section{Methodology}
%\subsection{Mathematical formulation}

The continuity and Navier-Stokes equations that describe the perturbed system, subject to small three-dimensional disturbances, are linearized and written
as

\begin{equation}
 \frac{\partial \widetilde{u}}{\partial x} + \frac{\partial
\widetilde{v}}{\partial y} + \frac{\partial
\widetilde{w}}{\partial z} = 0, \label{IVP2_continuity}
\end{equation}
\begin{equation}
 \frac{\partial \widetilde{u}}{\partial t} + U \frac{\partial
\widetilde{u}}{\partial x} + \widetilde{v} \frac{\partial
U}{\partial y} +  \frac{\partial \widetilde{p}}{\partial x} =
\frac{1}{Re} \nabla^2\widetilde{u}, \label{IVP2_NS1}
\end{equation}
\begin{equation}
 \frac{\partial \widetilde{v}}{\partial t} + U \frac{\partial
\widetilde{v}}{\partial x} + \frac{\partial
\widetilde{p}}{\partial y} = \frac{1}{Re} \nabla^2\widetilde{v},
\label{IVP2_NS2}
\end{equation}
\begin{equation}
 \frac{\partial \widetilde{w}}{\partial t} + U \frac{\partial
\widetilde{w}}{\partial x} + \frac{\partial
\widetilde{p}}{\partial z} = \frac{1}{Re} \nabla^2\widetilde{w},
\label{IVP2_NS3}
\end{equation}

\noindent where ($\widetilde{u}(x, y, z, t)$, $\widetilde{v}(x, y,z, t)$, $\widetilde{w}(x, y, z, t)$) and $\widetilde{p}(x, y, z,t)$ are the perturbation velocity components and pressure, respectively. $U$ and $dU/dy$ indicate the base flow profile (under the near-parallelism assumption) and its first derivative in the shear direction. For the channel flow, the independent spatial variable $z$ is defined from $-\infty$ to $+\infty$, the $x$ variable from $-\infty$ to $+\infty$, and the $y$ from $-1$ to $1$. For the wake flow, $z$ is defined from $-\infty$ to $+\infty$, $x$ from $0$ to $+\infty$, and $y$ from $-\infty$ to $+\infty$. All the physical quantities are normalized with respect to a typical velocity (the free stream velocity, $U_f$, and the centerline velocity, $U_0$, for the 2D wake and the plane Poiseuille flow, respectively), a characteristic length scale (the body diameter, $D$, and the channel half-width, $h$, for the 2D wake and the plane Poiseuille flow, respectively), and the kinematic viscosity, $\nu$.

The  base flow of the wake is approximated at an intermediate ($x_0=10$) and at a far longitudinal station ($x_0=50$), through two-dimensional analytical expansion solutions \cite{TB03} of the Navier-Stokes equations. Assuming that the bluff-body wake is a slowly  evolving spatial system, the base flow is frozen at each longitudinal station past the body, by using the first orders ($n = 0, 1$) of the expansion solutions \cite{TB03}, $U(y; x_0, Re)=1 - a x_0^{-1/2} \rm
exp\left(- \frac {Re}{ 4} \frac {y}{ x_0}\right)$,
%\begin{eqnarray}
%\nonumber U(y; x_0, Re)=1 - a x_0^{-1/2} \rm
%e^{\displaystyle{- \frac {Re}{ 4} \frac {y}{ x_0}}}, \label{IVP2_U_wake_profile}
%\end{eqnarray}
where $a$ is related to the drag coefficient $C_D$ ($a =
\textstyle{\frac 1 4} (Re / \pi)^{1/2} c_D(Re)$, see \cite{TB03}), and $x_0$ is the streamwise longitudinal station.

The plane channel flow is homogeneous in the $x$ direction and is represented by the Poiseuille solution, $U(y)=1-y^2$.

By combining equations (\ref{IVP2_continuity}) to (\ref{IVP2_NS3}) to eliminate the pressure terms, the perturbed system is fully described in terms of velocity and vorticity \cite{CD90} by the following equations:

\begin{eqnarray}
\nabla^2\widetilde{v} &=& \widetilde{\Gamma}, \\
\left (\frac{\partial }{\partial t} + U \frac{\partial }{\partial
x}\right)\widetilde{\Gamma} - \frac{\partial
\widetilde{v}}{\partial x}\frac{d^2U}{dy^2} &=& \frac{1}{Re}
\nabla^2 \widetilde{\Gamma}, \\
\left (\frac{\partial }{\partial t} + U \frac{\partial }{\partial
x}\right)\widetilde{\omega}_y + \frac{\partial \widetilde{v}}{\partial
z}\frac{dU}{dy} &=& \frac{1}{Re} \nabla^2 \widetilde{\omega}_y,
\label{IVP2_SQUIRE}
\end{eqnarray}

\noindent where $\widetilde{\omega}_y$ is the transversal
component of the perturbation vorticity. On the basis of kinematics, the following relation holds:
\begin{equation}
\widetilde{\Gamma} = \frac{\partial \widetilde{\omega}_z}{\partial x} - \frac{\partial \widetilde{\omega}_x}{\partial z}. \label{IVP2_Vel_Vor_Kin}
\end{equation}
\noindent The physical quantity, $\widetilde{\Gamma}$, physically links the perturbation vorticity in the $x$ and $z$ directions ($\widetilde{\omega}_x$ and $\widetilde{\omega}_z$, respectively) to the perturbation velocity field.

The perturbations are then Fourier transformed in the $x$ and $z$ directions for the channel flow. Two real wavenumbers, $\alpha$ and $\gamma$, are introduced along the $x$ and $z$ coordinates, respectively. A combined Laplace-Fourier decomposition is performed for the wake flow in the $x$ and $z$ directions. In this case, a complex wavenumber $\alpha = \alpha_r + i \alpha_i$ can be introduced along the $x$ coordinate, as well as a real wavenumber, $\gamma$, along the $z$ coordinate.
The perturbation quantities
$(\widetilde{v}, \widetilde{\Gamma}, \widetilde{\omega}_y)$
involved in the system dynamics are now indicated as $(\hat{v},
\hat{\Gamma}, \hat{\omega}_y)$, where

\begin{equation}\label{IVP2_Four_trans}
\hat{f}(y, t; \alpha, \gamma) = \int_{-\infty} ^{+\infty} \int_{-\infty}
^{+\infty} \widetilde{f}(x, y, z, t) e^{-i \alpha x -i \gamma z}
dx dz,
\end{equation}

\noindent indicates in the $\alpha - \gamma$ phase space the two-dimensional Fourier transform (in the case of the channel flow) of a general dependent variable, $\widetilde{f}$, and

\begin{equation}\label{IVP2_Four_trans}
\hat{g}(y, t; \alpha, \gamma) = \int_{-\infty} ^{+\infty} \int_{0}
^{+\infty} \widetilde{g}(x, y, z, t) e^{-i \alpha x -i \gamma z}
dx dz,
\end{equation}

\noindent indicates the two-dimensional Laplace-Fourier transform (in the case of the wake flow) of a general dependent variable, $\widetilde{g}$. To obtain a finite perturbation kinetic energy, the
imaginary part, $\alpha_i$, of the Laplace transformed complex longitudinal wavenumber
can only assume non-negative values and can thus be defined as a spatial damping rate in the  streamwise direction.
In so doing, perturbative
waves can spatially decay ($\alpha_i > 0$) or remain constant in
amplitude ($\alpha_i = 0$). Here, for the sake of simplicity, we have $\alpha_i=0$, therefore $\alpha=\alpha_r$. The governing partial
differential equations we consider are thus

\begin{eqnarray} \label{IVP2_fou1}
\frac{\partial^2 \hat{v}}{\partial y^2} &-& k^2 \hat{v} = \hat{\Gamma}, \\
\frac{\partial \hat{\Gamma}}{\partial t} &=& - i k cos(\phi) U \hat{\Gamma} + i k cos(\phi) \frac{d^2 U}{dy^2} \hat{v} + \frac{1}{Re} \left(\frac{\partial^2 \hat{\Gamma}}{\partial y^2} -
k^2 \hat{\Gamma}\right),\label{IVP2_fou2} \\
 \frac{\partial \hat{\omega}_{y}}{\partial t} &=& - i k cos(\phi) U \hat{\omega}_{y}
  - i k sin(\phi) \frac{dU}{dy} \hat{v} + \frac{1}{Re} \left(\frac{\partial^2 \hat{\omega}_y}{\partial
y^2} - k^2
\hat{\omega}_y\right),\label{IVP2_fou3}
\end{eqnarray}

\noindent where $\phi = tan^{-1}(\gamma/\alpha)$ is the perturbation obliquity angle with respect to the $x$-$y$ plane, $k = \sqrt{\alpha^2 + \gamma^2}$ is the polar wavenumber and $\alpha = k cos(\phi)$, $\gamma = k sin(\phi)$ are the wavenumber components in the $x$ and $z$ directions, respectively.

Various initial conditions can be used to explore the transient behavior. The important feature here is the ability to make arbitrary specifications. It is physically reasonable to assume that the natural issues affecting the initial conditions are the symmetry and the spatial lateral distribution of disturbances. It has been observed \cite{STC09,S08,CJLJ97} that, keeping all the other parameters fixed, if the perturbation
oscillates rapidly or mainly lies outside the shear region then, for a stable
configuration, the final damping is accelerated while, for an unstable configuration,
the asymptotic growth is delayed. However, the general qualitative scenario is not altered. Therefore, to perform a more synthetic perturbative analysis, we only focus on symmetric and asymmetric inputs which are localized and distributed over the whole shear region. The transversal vorticity $\hat{\omega}_y(y,t)$ is initially taken equal to zero to highlight the three-dimensionality net contribution on its temporal evolution. The effects of non-zero initial conditions on the transversal vorticity $\hat{\omega}_y(y,t)$ can be found in \cite{S08,CJLJ97}. The imposed initial conditions are reported in Table S1, for the channel and wake flows. $\Omega(\alpha, \gamma)$ is the phase space transform of the $x$-$z$ dependence prescribed at time $t=0$. Here, we set $\Omega(\alpha, \gamma)=1$, which means that no wavenumber is initially biased in the phase space.

\begin{table}
  \centering
\begin{tabular}{|c|c|c|}
  \hline
  % after \\: \hline or \cline{col1-col2} \cline{col3-col4} ...
   & Channel flow & Wake flow \\
   & & \\
  \hline
  & $\Omega(\alpha, \gamma)(1-y^2)^2$ & $\Omega(\alpha, \gamma)exp(-y^2)cos(y)$ \\
  $\hat{v}(y,t=0)$ & or & or \\
  & $\Omega(\alpha, \gamma)y(1-y^2)^2$ & $\Omega(\alpha, \gamma)exp(-y^2)sin(y)$\\
  \hline
  $\hat{\omega}_y(y,t=0)$ & 0 & 0 \\
  \hline
\end{tabular}
\label{table}
\caption{Initial conditions for the channel and wake flows.}
\end{table}

\noindent For the channel flow no-slip and impermeability boundary conditions are imposed,
\begin{equation}
\hat{v}(y=\pm1,t) = \frac{d\hat{v}}{dy}(y=\pm1,t) = \hat{\omega}_y(y=\pm1,t) = 0,
\end{equation}
\noindent while for the wake flow uniformity at infinity and finiteness of the energy are imposed,
\begin{equation}
\hat{v}(y\rightarrow\pm\infty,t) = \frac{d\hat{v}}{dy}(y\rightarrow\pm\infty,t) = \hat{\omega}_y(y\rightarrow\pm\infty,t) = 0.
\end{equation}

To measure the growth of the perturbations, we define the kinetic energy density, $e$,
\begin{eqnarray}
\nonumber e(t; \alpha, \gamma) &=& \frac{1}{2} \int_{-y_f}^{+y_f}
(|\hat{u}|^2
+ |\hat{v}|^2 + |\hat{w}|^2) dy,
%&=%& \frac{1}{2} \frac{1}{|\alpha^2 + \gamma^2|}\int_{-y_f}^{+y_f}
%\left(\left|\frac{\partial \hat{v}}{\partial y}\right|^2 + |\alpha^2 + \gamma^2|
%|\hat{v}|^2 + |\hat{\omega}_y |^2\right) dy,
\end{eqnarray}
\noindent where $-y_f$ and $y_f$ are the computational limits of the domain, while $\hat{u}$, $\hat{v}$ and $\hat{w}$ are the transformed velocity components of the perturbation field. We can also define the amplification factor, $G$, as the kinetic energy density normalized with respect to its initial value:
\begin{equation}
G(t; \alpha, \gamma) = \frac{e(t; \alpha, \gamma)}{e(t=0; \alpha,
\gamma)}.
\end{equation}

Assuming that the temporal asymptotic behavior of the linear perturbations is exponential, the temporal growth rate, $r$, can be defined as $r(t; \alpha, \gamma) = log(e)/(2t)$.
%\begin{equation}
%r(t; \alpha, \gamma) = \frac{\displaystyle{\frac{dG}{dt}}}{G}. \label{IVP2_tgr}
%\end{equation}

\noindent This quantity has a precise meaning when the asymptotic state is reached, that is, when it becomes a constant.

The angular frequency (pulsation), $\omega$, of the perturbation can be
defined considering the phase, $\varphi$, of the complex wave at a
fixed transversal station $y_0$ (for example, $y_0 = 1$ for the wake flow and $y_0 = 0.5$ for the channel flow)

\begin{eqnarray}
\varphi(t; \alpha, \gamma) = arg(\hat{v}(y=y_0, t; \alpha, \gamma))
= tan^{-1}(\frac{\hat{v}_i(y=y_0, t; \alpha, \gamma)}{\hat{v}_r(y=y_0,
t; \alpha, \gamma)}),
\end{eqnarray}
\noindent and then computing the time derivative of the phase
perturbation, $\varphi$,
\begin{eqnarray}
\omega(t; \alpha, \gamma) = \frac{|d \varphi(t; \alpha, \gamma)|}{dt}.
\end{eqnarray}

\noindent Although defined at any time $t$, the frequency $\omega$ is
here referred to as an asymptotic property of the perturbation.

Equations (\ref{IVP2_fou1})-(\ref{IVP2_fou3}) are numerically solved by the method of lines: the equations are first discretized in the spatial domain and then  integrated in time. The spatial derivatives in the $y$ domain are discretized using a second-order finite difference scheme  for the first and second derivatives. One-sided differences are adopted at the boundaries, while central differenced derivatives are used in the remaining part of the domain. The spatial grid is uniform with a spatial step, $h$, which is equal to 0.05 and 0.004, for the wake and channel flows, respectively. Since the wake flow is spatially unbounded in the transversal direction, the spatial domain, $[-y_f, y_f]$, is chosen so that the numerical solutions are insensitive to further extensions of the computational domain size. Equations (\ref{IVP2_fou1})-(\ref{IVP2_fou3}) are then integrated in time by means of an adaptative one-step solver, based on an explicit Runge-Kutta 3(2) formula.% and implemented in the \verb"ode23" Matlab function [S11, S12]. The choice of the \verb"ode23" routine is a good compromise between nonstiff solvers (\verb"ode45" and \verb"ode113"), which give a higher order of accuracy, and stiff solvers (\verb"ode15s" and \verb"ode23s"), which can in general be more efficient.
\section{Database setting}

%The set of solutions of the linearized Navier-Stokes perturbative equations is accessible at the same web site where the Matlab scripts can be found (\url{https://130.192.25.166}, username: \verb"guest", password: \verb"etipso" -- in case of publication the database will be moved to a freely accessible web repository). In following, the database structure is described.
The Matlab scripts and the set of solutions of the linearized Navier-Stokes perturbative equations can be found at \rm{https://130.192.25.166} (username: \verb"guest", password: \verb"etipso"). By using the "File Browser" it is possible to access the disks

\noindent \verb"/LaCie(usb)#2/Linearized_NS_solutions"

For the channel flow, the perturbative analysis considers 4 parameters (Reynolds number, symmetry/asymmetry, angle of obliquity and wavenumber) and 660 solutions are obtained. If one, by using the "File Browser", goes to

\noindent \verb"/LaCie(usb)#2/Linearized_NS_solutions/channel",

\noindent he will find the database organized as in Fig. S4. Every folder, corresponding to a certain specification of the above parameters, contains the following text files:

\begin{itemize}
\item \verb"prefix_t_n.txt": the temporal points, $M$, at which the solutions are computed through the Matlab code, are reported in column;
\item \verb"prefix_u_n.txt", \verb"prefix_v_n.txt", \verb"prefix_w_n.txt": these files contain the perturbation velocity field components. Each of them has two columns, the first one for the real part and the second one for the imaginary part of the velocity component. The column length is $MxN$, where $N$ is the number of spatial grid points ($N=501$ for the channel flow) and $M$ are the temporal instants. For every fixed time, the velocity spatial distributions are put in column;
\item \verb"prefix_omega_y_n.txt": this file contains the transversal vorticity component and is structured as the above velocity field files;
\item \verb"prefix_energy_n.txt": in this file, the kinetic energy density, $e$, is reported. The first column expresses the temporal points, $M$, and the second one the corresponding energy values.
\end{itemize}

\begin{figure}
\centering
  \includegraphics[width=\columnwidth]{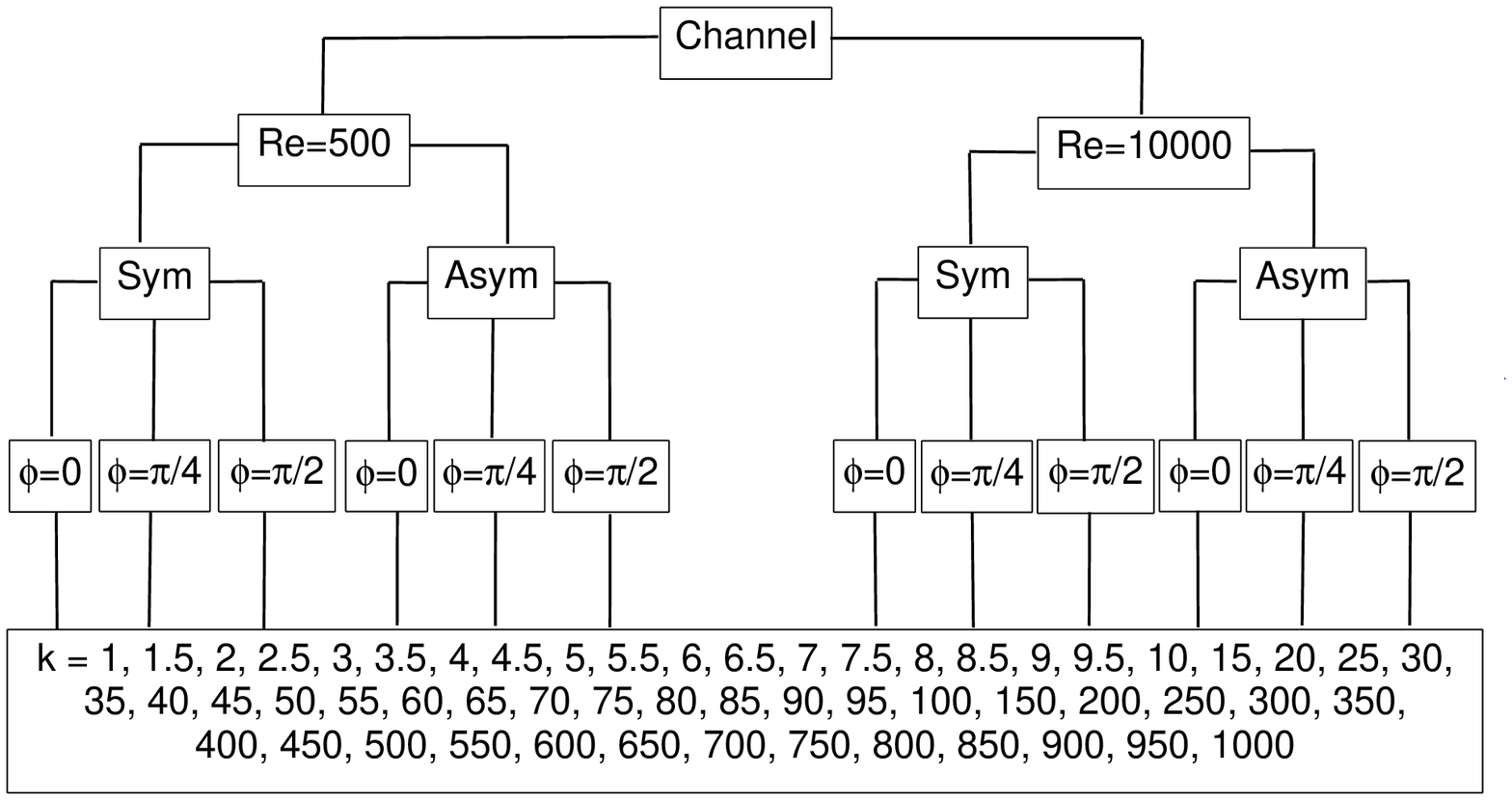}
  \label{fig:canale}
  \caption{Database structure for the channel flow.}
\end{figure}

\begin{figure}
\centering
  \includegraphics[width=\columnwidth]{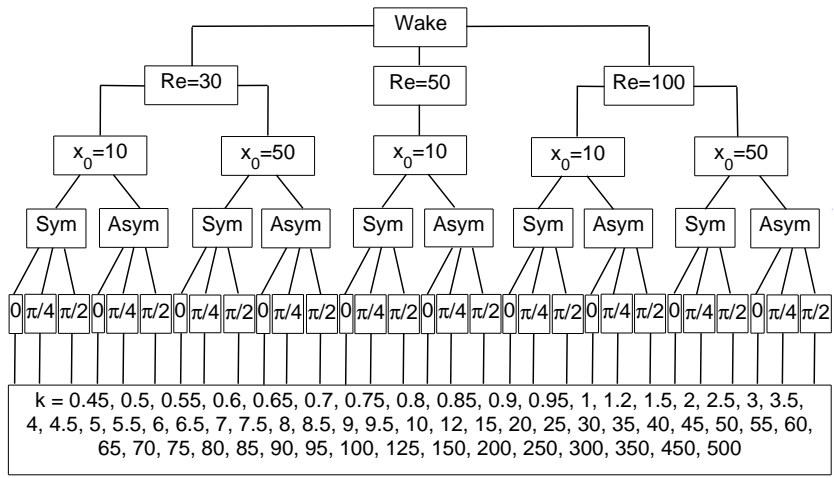}
  \label{fig:scia}
  \caption{Database structure for the wake flow.}
\end{figure}

\noindent Each of these files contains, as a prefix in its name, the parameter information and has an increasing number, $n$, as a suffix. This integer number, $n$, accounts for the fact that outputs are periodically saved (after a variable temporal interval) as the equations are integrated in time.

Let us make an example. Suppose we are interested in the simulation with parameters $Re=500$, symmetric initial conditions, $\phi=\pi/4$, $k=3$. If one goes to

\noindent \verb"/channel/Re_500/Re_500_sym/Re_500_sym_phi_45/Re_500_sym_phi_45_k_3",

\noindent he will find the above text files with prefix \verb"Re_500_sym_phi_45_k_3" and suffix $n\geq1$.

For the wake flow, data are organized in an analogous way to the one described for the channel flow and can be found in the following folder,

\noindent \verb"/LaCie(usb)#2/Linearized_NS_solutions/wake".

\noindent It should be recalled that 5 parameters are here considered (Reynolds number, wake position, symmetry/asymmetry, angle of obliquity and wavenumber) and 1740 solutions are computed (see Fig. S5, where the scheme of the database structure for the wake flow is shown). Therefore, in the corresponding folders and files, as part of the suffix, information on the chosen downstream station, $x_0$, are added similarly to what done for the other parameters. Moreover, concerning the perturbation velocity and the transversal vorticity files, it should be mentioned that the spatial domain is enlarged when long perturbative waves are analyzed. Therefore, the number of grid points, $N$, depends on the wavenumber considered: $N=801$ if $k>1$, $N=1201$ if $k\in[0.75, 1]$, $N=1601$ if $k\in[0.45, 0.7]$.